\documentclass[aps,prb,twocolumn,superscriptaddress,showkeys,longbibliography,nobibnotes]{revtex4-2}
\usepackage[colorlinks=true,linkcolor=blue,urlcolor=blue,citecolor=blue,anchorcolor=blue]{hyperref}
\usepackage{textcomp,gensymb}
\usepackage{graphicx}
\usepackage[T1]{fontenc}
\usepackage{txfonts}
\usepackage{array}
\usepackage{float}
\usepackage{lipsum}
\usepackage{subcaption}
\usepackage{booktabs}

\begin{document}
	
\title{Understanding the evolution of the magnetic ground state in Ba$_4$NaRu$_3$O$_{12}$}

\author{Shruti Chakravarty}
\affiliation{Department of Physics, Indian Institute of Science Education and Research, Pune, India}

\author{Pascal Manuel}
\affiliation{ISIS Pulsed Neutron Source, STFC Rutherford Appleton Laboratory, Didcot, Oxfordshire OX11 0QX, United Kingdom}

\author{Antonio Cervellino}
\affiliation{Swiss Light Source, Paul Scherrer Institute, CH-5232 Villigen, Switzerland}

\author{Sunil Nair}
\email[Corresponding author: ]{sunil@iiserpune.ac.in}
\affiliation{Department of Physics, Indian Institute of Science Education and Research, Pune, India}

\date{\today}

\begin{abstract}

We report a comprehensive investigation of the quadruple perovskite Ba$_4$NaRu$_3$O$_{12}$, in which we discover a robust spin-lattice coupled ground state characterized by a long-range antiferromagnetic ordering at $T_N \sim$ 257 K. The system's unique structural motif of three symmetrically distinct magnetic ions, including Ru dimers separated by non-magnetic layers, is intimately correlated with its magnetic behaviour, as evidenced by temperature-dependent diffraction measurements and specific heat data. The powder neutron diffraction patterns at 13 K showed that the spins within the dimers are antiparallel, leading to a net zero moment contribution and a staggered arrangement of the triangular layers formed by the Ru moments within the corner-shared octahedra along the $c$-axis. The low-temperature specific heat revealed an extra boson peak contribution from optical modes with a maximum vibrational energy of $\sim$55cm$^{-1}$. The charge transport exhibited variable-range hopping (VRH) behaviour below $T_N$, with a stronger energy-dependence than expected from the Efros-Shklovskii model, suggesting the presence of multiparticle correlation effects.

\end{abstract}


\maketitle

\section{Introduction}

Perovskites are one of the most celebrated material classes in condensed matter physics due to their unprecedented compositional flexibility and versatility. The ideal perovskite structure, a neatly arranged octahedral network, can be tweaked to generate a variety of complexity, giving rise to genuinely intriguing physics. Substitution of A- or B-sites of the parent ABO$_3$ structure produces various kinds of layerings of which the hexagonal layered systems are of special interest due to their inherent geometric frustration. Various 4d and 5d double- (1:1 substitution of the B-site) and triple- (1:2 substitution of the B-site) perovskite systems are known to host a plethora of exotic magnetic states \cite{Kimber2012,Ziat2017,chen2020,Terasaki2017}.\\

A 1:3 substitution of the A-site generates highly anisotropic crystal systems with very interesting material properties. These systems have recently gained traction due to their excellent dielectric \cite{ahmadipour2016short} and catalytic properties \cite{liu2022}. They are categorized as distorted cubic perovskites and their structures are strongly influenced by the type of Glazer tilt system \cite{glazer1972} involved. For example, large $a^+a^+a^+$ tilts produce the AA'$_3$B$_4$O$_{12}$ structure where the strong tilting of the B octahedra forces the A' sites to settle in a square planar geometry. On the other hand, strong tilting in the $a^+a^+c^-$ directions generates the A$_2$A'A''B$_4$O$_{12}$ structure possessing an inherent columnar-order\cite{belik2022triple}. These materials, although with a huge potential for the exploration of symmetry-influenced physical properties, require the use of simultaneous high-pressure and high-temperature environments to be stabilized.\\

The B-site substituted systems, conversely, show very interesting \emph{quasimolecular} physics arising from the cation dimers and trimers governing their magnetism. For example, in systems like Ba$_3$NbRu$_3$O$_{12}$ \cite{nguyen2018} and Ba$_4$LnM$_3$O$_{12}$ \cite{shimoda2010magnetic,shimoda2008synthesis} (M = Ru, Ir), Ru/Ir trimers isolated by nonmagnetic layers show fascinating frustration-based physics with either long-range order developing at extremely low temperatures or not at all. The relative proximity of the transition-metal (TM) cations within these trimers leads to the formation of molecular orbitals which dictate the physical properties in these systems. The atomic $d$-orbitals of two relatively close TM cations, in a local $D_{3h}$ symmetry, linearly combine to produce bonding and antibonding molecular orbitals. These states are strongly influenced by the geometry of the local field and the electron hopping probabilities between them. This makes $d$-orbital compounds an attractive platform to study instabilities leading to quantum phase transitions\cite{yuan2024highly}, due to their ability for direct exchange and high degree of tunability. Additionally, compounds containing isolated groups of magnetic ions as repeating motifs have often surprised physicists with their exotic properties and complex magnetic ground states\cite{basu2020,Kimber2012} despite disruptions in the exchange pathway. This highlights the importance of next-nearest neighbour superexchange mediated interactions in these materials.\\

Ba$_4$NaRu$_3$O$_{12}$ uniquely stands out in this family of B-site substituted quadruple perovskites in that it is neither a network of isolated or connected Ru/Ir trimers like Ba$_5$Ru$_3$O$_{12}$ \cite{basu2020} and Ba$_4$(Ru/Ir)$_3$O$_{10}$ \cite{klein2011,Cao2020} nor of isolated  Ru dimers like its sister triple-perovskite Ba$_3$NaRu$_2$O$_9$\cite{Kimber2012}. The repeating motif of this structure is a set of one RuO$_6$ octahedra connected via a corner shared oxygen to a Ru$_2$O$_9$ dimer separated by layers of nonmagnetic NaO$_6$ octahedra. To the best of our knowledge, no other layered ruthenate is known to possess such a motif. This material has also been scarcely studied in the literature, with few reports on it from nearly 20 years ago \cite{battle1992,kim1998,stitzer2003}, leaving much to be explored, especially its complex magnetism at low $T$. This motivates us to investigate this compound in-depth and explore its ground state's evolution as a function of temperature and characterize its correlation with lattice and charge degrees of freedom.\\

\begin{figure*}
	\centering
	\includegraphics[width=\textwidth]{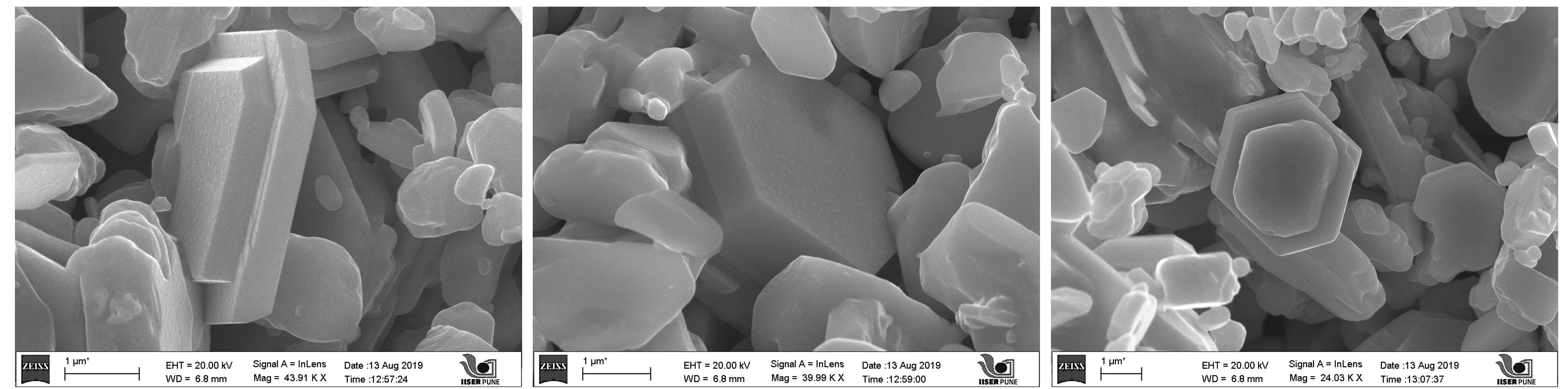}
	\caption{Scanning electron micrographs of Ba$_4$NaRu$_3$O$_{12}$ powder samples, taken at 20keV energy, capture the micron-sized hexagonal crystallites displaying a plate-like morphology.}
	\label{fesem}
\end{figure*}

\section{Experimental Methods}
Polycrystalline samples of Ba$_4$NaRu$_3$O$_{12}$ were prepared via solid state synthesis method using stoichiometric amounts of raw materials (BaCO$_{3}$, Na$_{2}$CO$_{3}$ and RuO$_{2}$) following the procedure detailed by Battle et al. \cite{battle1992}. To compensate for the volatility of Na, a 15\% excess of Na$_{2}$CO$_{3}$ was used and RuO$_2$ was heated for 2 - 3 hours at 250\textdegree C before weighing to remove any moisture. The precursors were mixed and ground in a mortar using ethanol until a fine, homogeneous mixture was obtained. The entire mixture was then pelletized to further reduce the loss of Na, placed in an alumina crucible and heated at 800\textdegree C for 12 hours in air. Hard, black pellets obtained at the end of the reaction were crushed and checked for phase purity through laboratory x-ray diffraction (XRD) using the {\footnotesize BRUKER D8 ADVANCE DIFFRACTOMETER} (CuK$_{\alpha}$ $\lambda$=1.5406\AA).
Elemental compositions and their homogeneity was also confirmed by using an energy dispersive X-ray spectrometer (EDS) (Ziess Ultra Plus). The average stoichiometry calculated from the statistical analysis of the EDS data is Ba$_{4}$Na$_{0.87\pm0.1}$Ru$_{3.02\pm0.2}$O$_{12}$, confirming the homogeneity and single phase nature of the sample. Field emission scanning electron microcopy (FESEM) images of hexagonal microstructure are shown in Fig.\ref{fesem}

Synchrotron x-ray powder diffraction (SXRD) was performed at the Swiss Light Source (PSI) on the ADDAMS beamline \cite{sls-mspd}. The wavelength of the x-ray was calculated by refining LaB$_6$ standard to be $\lambda = $ 0.49343 \AA{} and an instrument-resolution file (IRF) file was created (FWHM (\textdegree) vs 2$\theta$). Using this IRF file, the profile parameters were fixed and only microstructural and atomic parameters were refined for diffraction patterns collected at each $T = 5 - 300$ K. This allowed us to estimate changes in maximum strain (\%) with temperature. Rietveld refinement of the diffraction patterns were performed using the {\footnotesize FULLPROF SUITE} \cite{fullprof}.  Crystal Structures were generated using {\footnotesize VESTA} \cite{vesta}. Neutron time-of-flight (\textit{tof}) measurements were performed at the WISH diffractometer \cite{chapon2011wish} of ISIS Pulsed Neutron and Muon source in the $T$ range 13 - 300 K. Data collected on 5 banks were simultaneously refined using {\footnotesize FULLPROF} and Bank 4-7 (2$\theta=$121.66\textdegree) was chosen to be the pattern of choice with the most appropriate $q-$range and resolution for plotting. Specific Heat and Magnetization measurements were performed in a Quantum Design PPMS Dynacool. Resistivity measurements were performed in a Cryogenic Limited Cryogen Free Measurement System (CFMS) using a VTI-based resistivity option. 

\section{Results and Discussion}

\begin{figure}[h]
	\centering
	\includegraphics[width=\columnwidth]{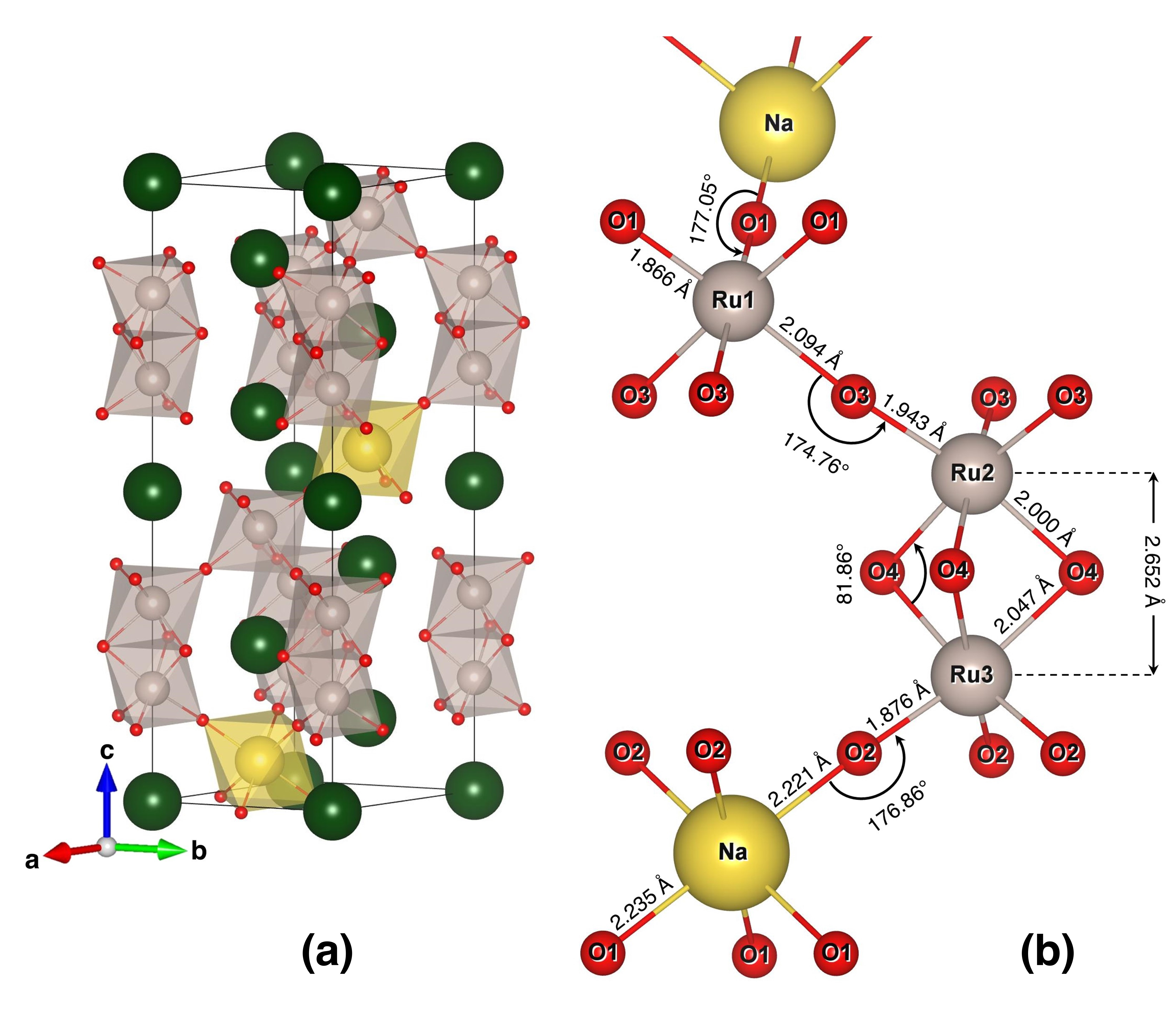}
	\includegraphics[width=\columnwidth]{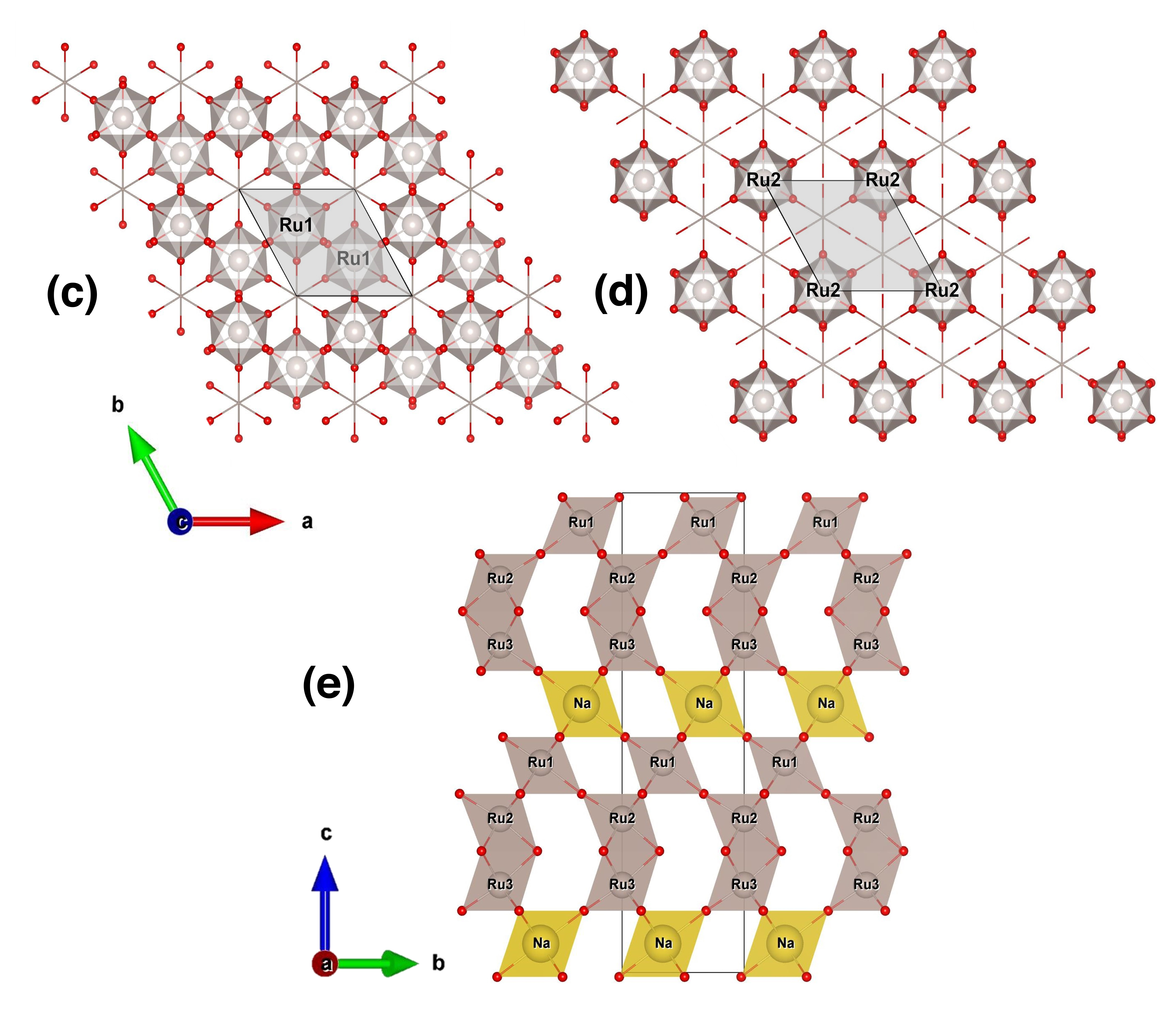}
	\caption{(a) Crystal structure of Ba$_4$NaRu$_3$O$_{12}$. (b) Bond lengths and angles of the exchange path calculated from at room temperature neutron diffraction data. The triangular network formed in the ab-plane by (c) Ru1/Na and (d) Ru2/Ru3 ions viewed along the c-axis. (e) The 8-layered unit cell viewed along the a-axis.}
	\label{structure}
\end{figure}

Ba$_4$NaRu$_3$O$_{12}$ crystallizes in a rare 8-layered hexagonal structure described by the non-centrosymmetric space group $P6_3mc$ (No. 186) as shown in Fig.\ref{structure}, with lattice parameters $a=b=$ 5.797(2) \AA{} and $c=$ 19.211(6) \AA{} .  The Ba ions have 4 crystallographically distinct sites and occupy corner and edges of the unit-cell as well as the voids between the octahedra. The Ru ions have three unique positions as well, two of them (Ru2, Ru3) forming a face-shared octahedra located on the unit cell edges, while one (Ru1) sits at the centre of an octahedra corner-linking the Ru$_2$O$_9$ and the NaO$_6$ octahedra. This structure is visually presented in Fig.\ref{structure}(a). Viewed along the $a$-axis the 8-layers (2 formula units, so $Z=2$) forming the unit cell can be seen in Fig.\ref{structure} (d). Fig.\ref{structure}(b) and (c) show the Ru1(/Na) and Ru2(/Ru3) layers in the $ab$-plane, highlighting the triangular arrangement of Ru cations in each layer. The structural parameters obtained from the Rietveld refinement of this structure are tabulated in Tables\ref{strucparNa}. The values obtained are consistent with previous reports \cite{battle1992,kim1998}.

\begin{figure}[h]
	\includegraphics[width=\columnwidth]{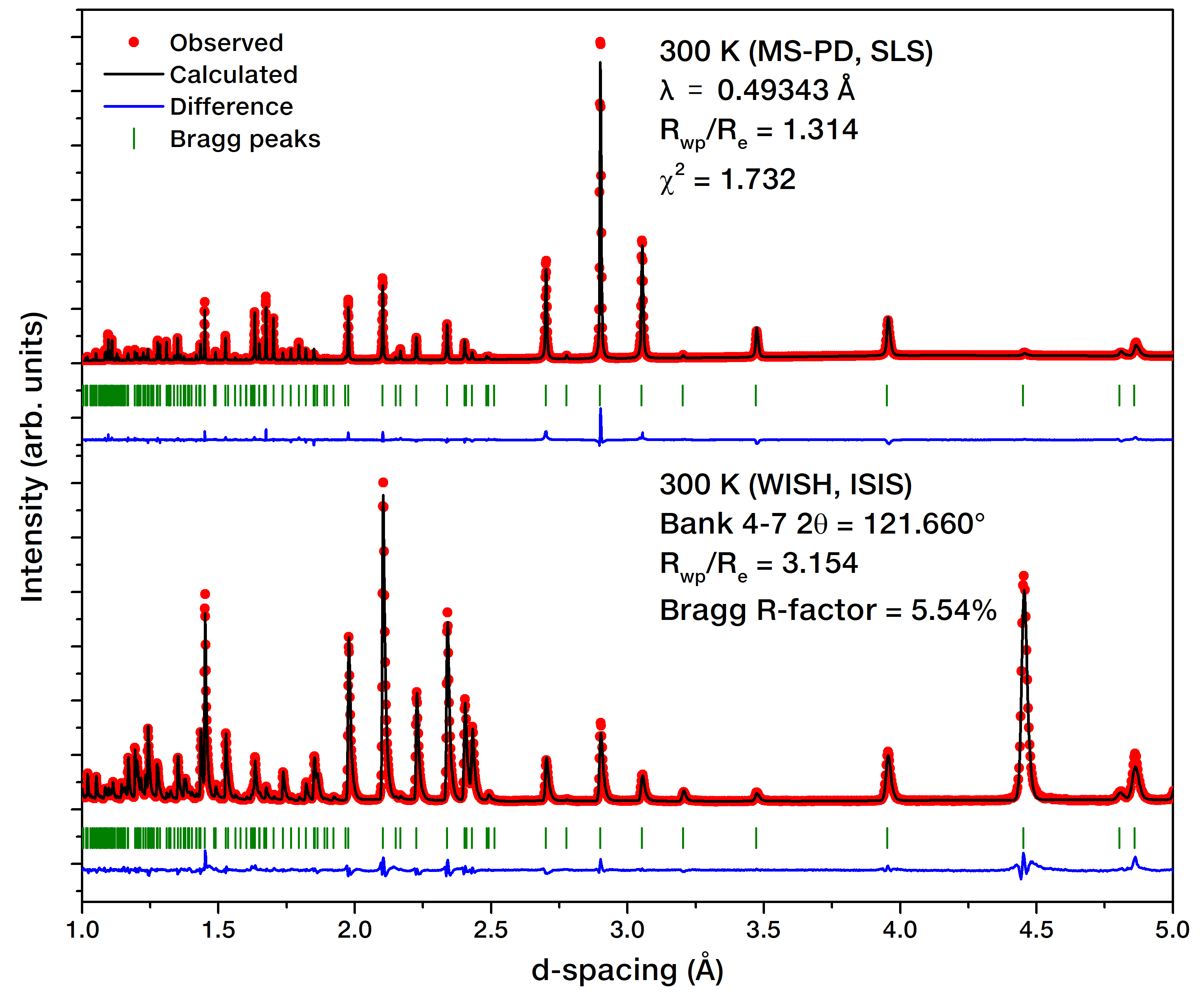}
	\caption{Room temperature neutron and synchrotron XRD powder diffraction patterns of Ba$_4$NaRu$_3$O$_{12}$ refined using {\footnotesize FULLPROF SUITE} \cite{fullprof}}
	\label{rtxnpd}
\end{figure}

Since the ionic radii of Na\textsuperscript{+} (1.02 \AA\cite{shannon1976}) and Ru\textsuperscript{5+} (0.565 \AA\cite{shannon1976}) vastly differ, no site disorder is expected and all ions fully occupy their sites. The refinement results confirm this too and return a total cell content of Ba$_8$Na$_2$Ru$_6$O$_{23.99}$. However, since Ba2, Ba3, Ba4, Na and Ru1 - all lie on the same plane [Wyckoff site: $2b$ (1/3,2/3,$z$)], the positions and thermal parameters required constraints to prevent the fit from diverging. The three Ru sites are inequivalent, but have the same site multiplicity. Fig.\ref{rtxnpd} shows the fit to the structural model in Table\ref{strucparNa} for synchrotron x-ray (top panel) and neutron (bottom panel) powder diffraction data at room temperature. The goodness of fits obtained are reasonable and confirm the accuracy of the model.

The distance between the Ru ions within the Ru$_2$O$_9$ dimer is 2.652 \AA{}, shorter than the Ru-Ru metallic bond length (2.6725\AA{} \cite{Jain2013}), suggesting a strong antiferromagnetic (AFM) direct interaction between Ru2 and Ru3. The observed average Ru---O bond length $\approx$1.97 \AA{} is very close to that expected  for an Ru\textsuperscript{5+}---O\textsuperscript{2-} bond (1.964\AA\cite{Gagne2020}), confirming the pentavalent oxidation state of Ru. The Ru octahedra are also slightly distorted as evidenced by the difference in the distances of the two uniquely located O\textsuperscript{2-} ions forming each of the Ru octahedra (Ru1---O1 = 1.866\AA, Ru1---O3 = 2.094\AA, Ru2---O3 = 1.943\AA, Ru2---O4 = 2.0\AA, Ru3---O4 = 2.05\AA, Ru3---O2 = 1.876\AA). Ru2 and Ru3 are both farther away from the O\textsuperscript{2-} connecting them (O4) than those on the opposite side (O3, O2 respectively) due to Coulomb repulsion. Due to the same reason, the Ru2---O3 bond is longer than the Ru3---O2 bond which experiences repulsion from Na\textsuperscript{+} instead of the more electropositive Ru\textsuperscript{5+}.  Distortion is present for the NaO$_6$ octahedra as well, likely due to its large size. 

The Ru2---O4---Ru3 superexchange  ($\angle_{ROR} = $ 81.86\textdegree) competes with the direct exchange by favoring a weak ferromagnetic (FM) coupling. There will be an AFM exchange across the Ru1---Ru2 path as well ($\angle_{ROR} = $ 174.76\textdegree) due to the half filled $t^3_{2g}$ of Ru\textsuperscript{5+}. These three magnetic ions are separated by non-magnetic NaO$_6$ layers. Thus, for the exchange path to extend throughout the crystal, a robust next-nearest-neighbour (NNN) exchange between Ru1 and Ru3 is imperative - most likely a super-superexchange interaction mediated by the NaO$_6$ anions. 

\begin{table}
	\centering
	Space Group: P6$_{3}$mc (Hexagonal, No. 186) \\
	a = b = 5.797(2)\AA{} , c = 19.211(6)\AA{}, $\alpha$ = $\beta$ = 90\degree , $\gamma$ = 120\degree \\
	
	\begin{ruledtabular}
		\begin{tabular}{cccccc}
			\textbf{Atom} & \textbf{Wyckoff site} & \textbf{x} & \textbf{y} & \textbf{z} & \textbf{B$_{iso}$} \\
			\midrule	
			Ba1 & 2a & 0      & 0      & 0       	      & 0.8(3)\\
			Ba2 & 2b & 1/3    & 2/3    & 0.123(33)        & 0.8(3)\\
			Ba3 & 2b & 1/3    & 2/3    & 0.371(03)        & 0.8(3)\\
			Ba4 & 2b & 1/3    & 2/3    & 0.746(93)        & 0.8(3)\\
			Na  & 2b & 1/3    & 2/3    & 0.559(93)        & 0.8(3)\\ 
			Ru1 & 2b & 1/3    & 2/3    & 0.937(03)        & 0.8(3)\\
			Ru2 & 2a & 0      & 0      & 0.320(31)        & 0.8(3)\\
			Ru3 & 2a & 0      & 0      & 0.182(41)  	  & 0.8(3)\\
			O1  & 6c & 0.488(03) & 0.511(97) & -0.009(27) & 1.0(2)\\
			O2  & 6c & 0.844(21) & 0.155(79) & 0.128(71)  & 1.0(2)\\
			O3  & 6c & 0.834(51) & 0.165(59) & 0.372(51)  & 1.0(2)\\
			O4  & 6c & 0.847(91) & 0.152(09) & 0.753(21)  & 1.0(2)\\
		\end{tabular}
	\end{ruledtabular}
	\caption{Room temperature structural parameters calculated from synchrotron x-ray and neutron powder diffraction}
	\label{strucparNa}
\end{table}

DC magnetic susceptibility measurements were performed under an applied field of $H=$ 1 kOe in the zero field-cooled (ZFC), field-cooled cooling (FCC) and field-cooled warming (FCW) modes. A clear cusp at $\sim$265 K marks the onset of long-range antiferromagnetic order. A significant hysteresis is seen between the magnetization measured during heating and cooling, suggesting the possibility of the transition being of the first-order. Note that the triple-perovskite version of this compound, Ba$_3$NaRu$_2$O$_9$, with Ru$_2$O$_9$ as the physically meaningful units, exhibits a first-order charge-ordering transition at 210 K, stabilized by a combination of single-ion and molecular degrees of freedom, and accompanied by a $P6_3/mmc\rightarrow P2/c$ symmetry transformation\cite{Kimber2012,Stitzer2002}. The N\'eel temperature, $T_N =$ 257.42 $\pm$ 0.07 K is calculated by plotting the derivative of the ZFC susceptibility with respect to temperature. Several subtle features can be seen below $T_N$, consistent with previous reports. Although, qualitatively the $M(T)$ curves look different, both Battle et al. \cite{battle1992} and Kim et al. \cite{kim1998} report anomalies in the magnetic susceptibility below $T_N$. The former reported a $\chi_{\text{ZFC}}$ showing an irreversibility below $T_N$, a plateau between $\sim 200-150$ K, and another broad peak centered around $\sim 110$ K. Kim et al. \cite{kim1998}, had performed a magnetic field-dependence analysis and came to the conclusion that all the features except for $T_N$ are suppressed at some critical field.\\ 

\begin{figure}[ht]
	\centering
	\includegraphics[width=\columnwidth]{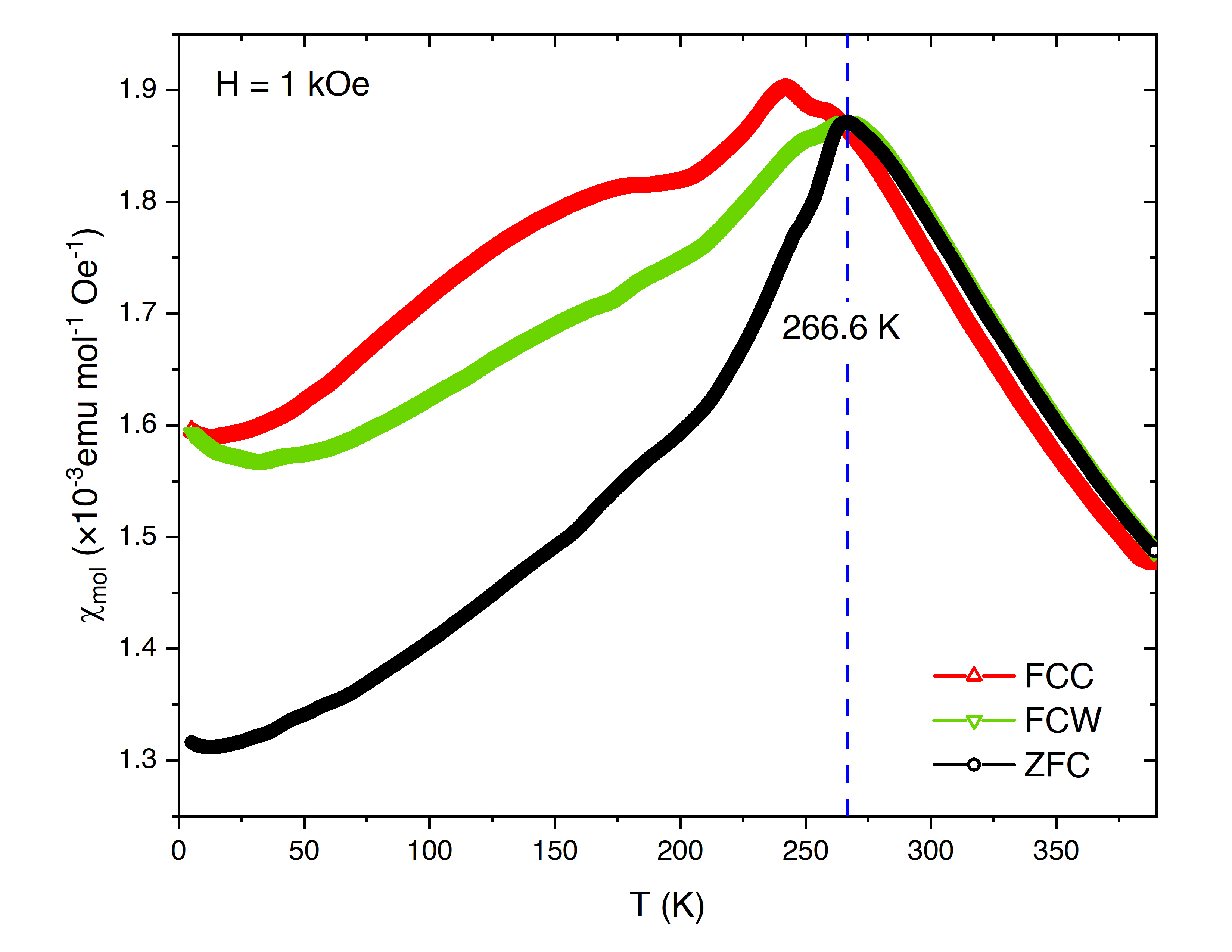}
	\caption{DC molar susceptibility $\chi_{mol}$ measured as a function of $T$ in ZFC, FCC and FCW protocols showing the antiferromagnetic transition at $T_N \sim$ 265 K along with other subtler features within the AFM phase.}
	\label{MTNa}
\end{figure}

\begin{figure}[ht]
	\centering
	\includegraphics[width=\columnwidth]{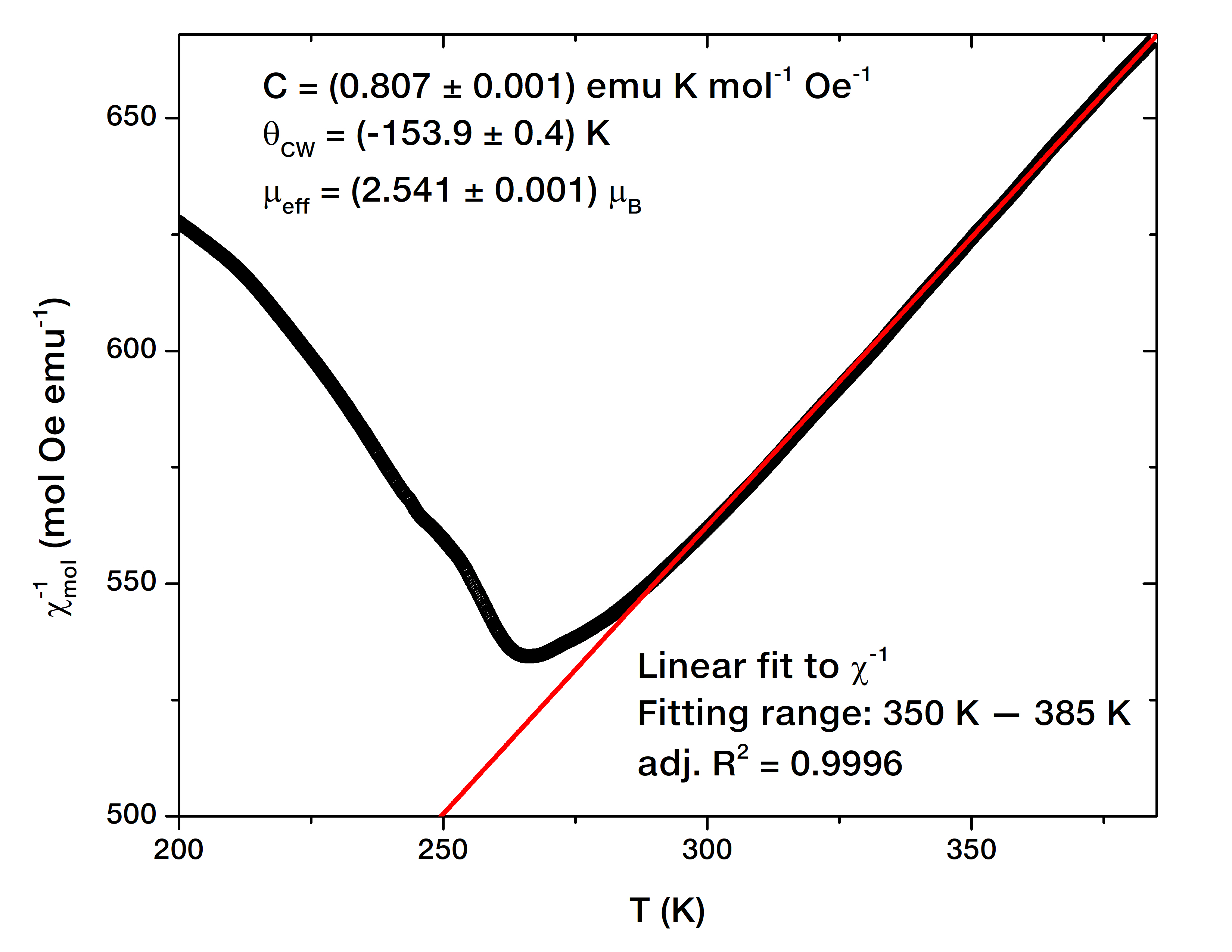}
	\caption{Curie-Weiss fit to the inverse magnetic susceptibility ($\chi_{mol}^{-1}(T)$) in the high $T$ range 350-385 K. The adj. R$^2$ value attests to the quality of the fit. $\theta_{CW}$ obtained from the fit demonstrates the dominance of antiferromagnetic interaction and the Ru moment per formula unit is suppressed to a value of $\sim$2.5$\mu_B$.}
	\label{1/chi}
\end{figure}

The inverse molar susceptibility $1/\chi_{mol}$ as a function of temperature $T$ is shown in Fig.\ref{1/chi}(b) showing Curie-Weiss (CW) behaviour down to $\sim$ 290 K, below which the system enters the transition region. To ensure the correct estimation of  CW parameters, the fitting has been performed across the highest measured $T$ range between 350 - 385 K (388 points). The Curie constant was calculated to be $C = 0.8$ emu K mol$^{-1}$Oe$^{-1}$ and the CW temperature $\theta_{CW} = -153.9 \pm 0.4$ K confirming the dominant antiferromagnetic correlations. 
Kim et al., \cite{kim1998} had performed high $T$ magnetization measurements on this material and shown that above 520 K, a CW fit to $\chi^{-1}$ yields a $\theta_{CW}=$ -1332 K and an effective magnetic moment, $\mu_{eff}=$ 3.74 $\mu_B$, nearly equal to the spin-only moment expected for a Ru$^{5+}$ $S=3/2$ state ($\mu_{so} = 2\sqrt{S(S+1)}  = 3.87 \mu_B$). They noted a change of slope in $\chi^{-1}$ below $\sim$520 K resulting in a significant reduction in $\theta_{CW} =$ 221 K and $\mu_{eff} = $ 2.32 $\mu_B$. This is attributed to the development of AFM correlations within the Ru$_2$O$_9$ dimers which lowers the measured bulk magnetization. Since our measurements were limited to 380 K, we likely probe this already correlated paramagnetic phase and calculate the (reduced) $\mu_{eff}$ per formula unit to be 2.541 $\pm$ 0.001 $\mu_B$. 

\begin{figure*}[ht]
	\centering
	\includegraphics[width=0.9\textwidth]{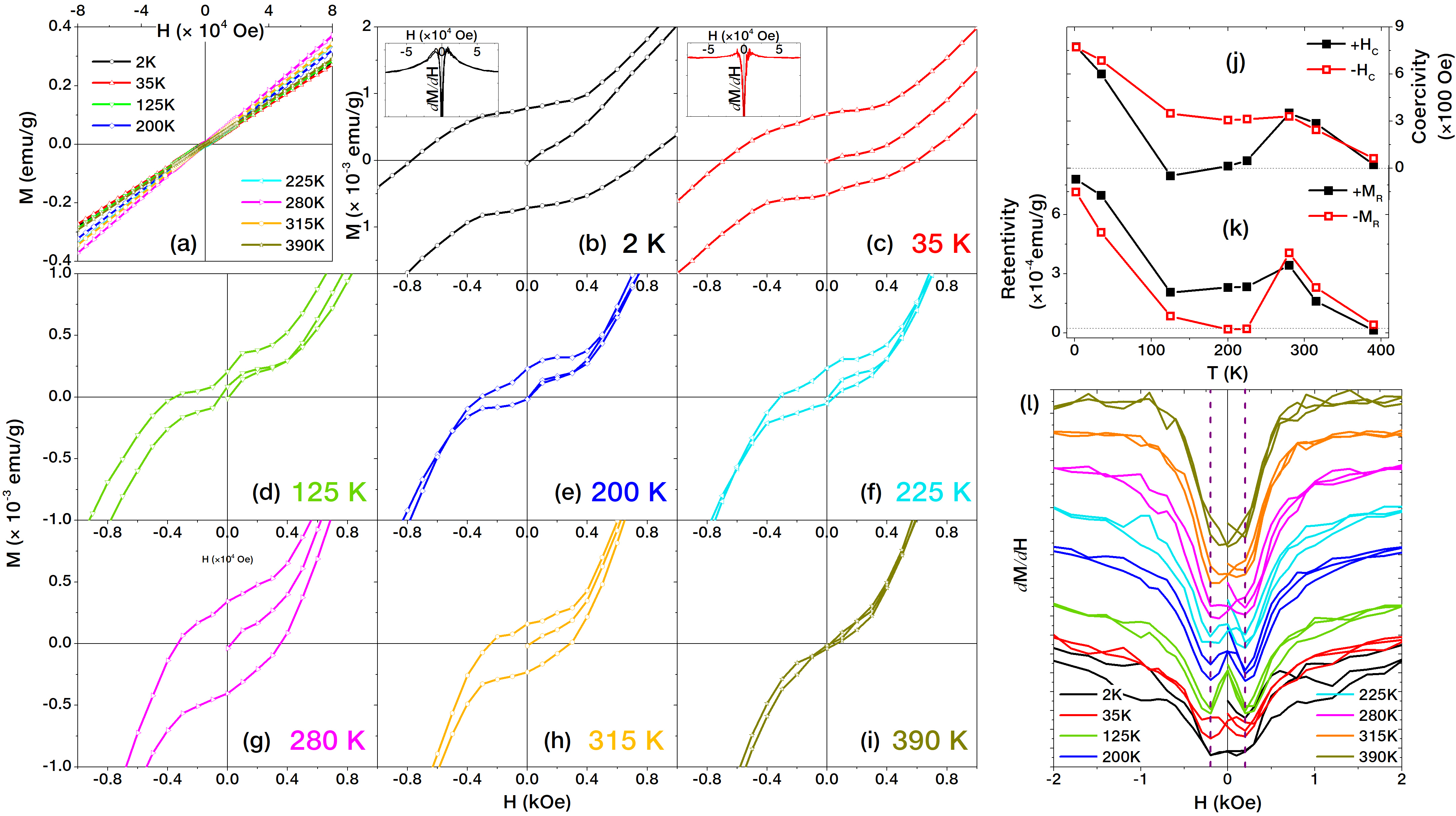}
	\caption{(a) $M(H)$ isotherms measured in the range [-8 T, 8T] for 8 temperatures between 2 - 400 K, (b) - (i) low field regions showing a finite loop and low-field features, insets of (b) and (c) show d$M$/d$H$ in the $H$ range [-8 T, 8 T] and (l) for $H \in$ [-0.1  T, 0.1 T], (j) and (k) show the coercivity ($H_c$) and Retentivity (or Remnant Magnetization, $M_R$) as function of temperature.}
	\label{MHNa}
\end{figure*}

\begin{figure}[h]
	\centering
	\includegraphics[width=\columnwidth]{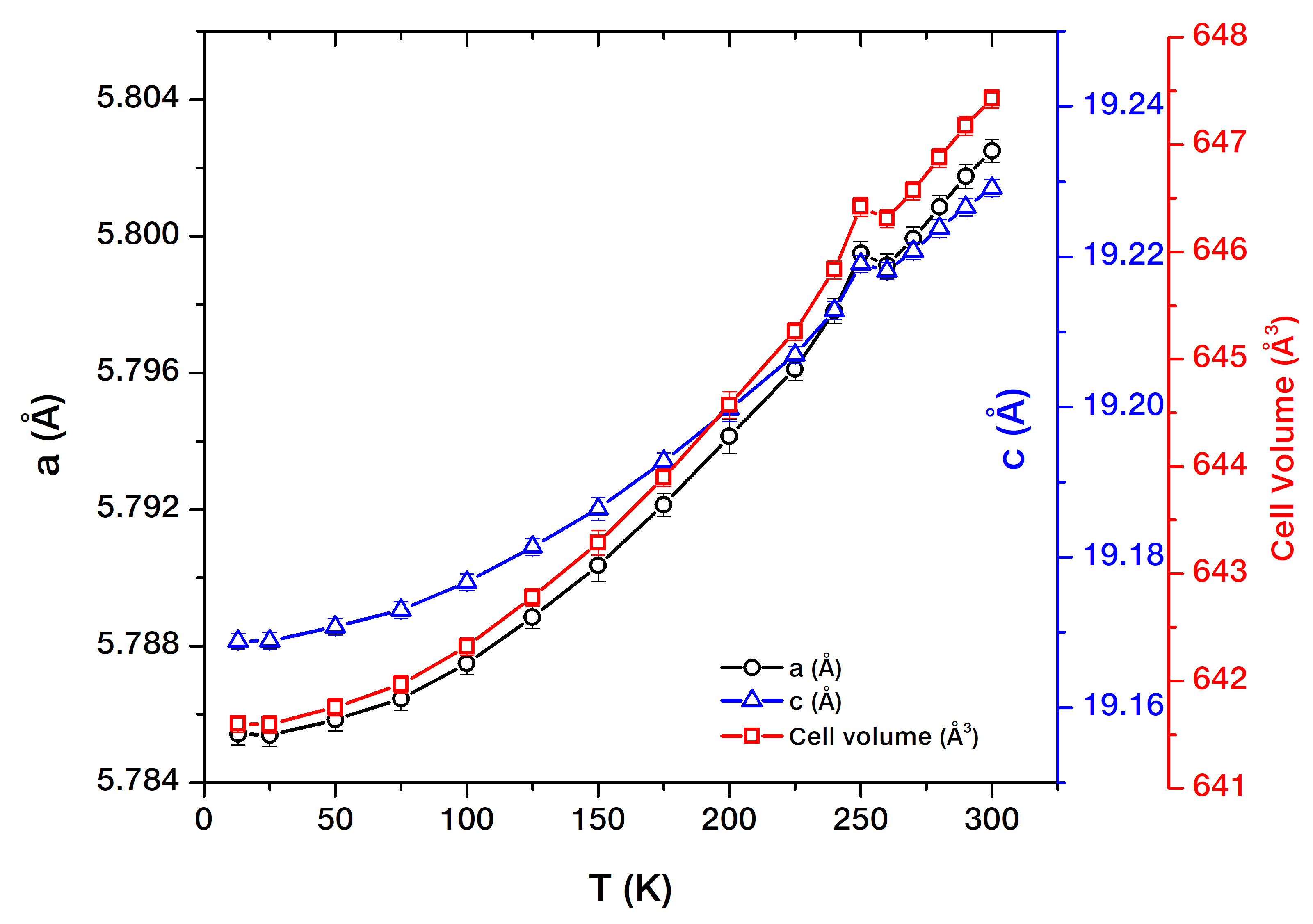}
	\caption{Evolution of lattice constants $a$, $c$ and the unit cell volume as a function of temperature ($T$), obtained from $T-$dependendent NPD. The antiferromagnetic transition is marked by the feature at 250 K. The error bars are small and marked on the graph.}
	\label{Tdepuc}
\end{figure}

\begin{figure}[h]
	\centering
	\includegraphics[width=0.9\columnwidth]{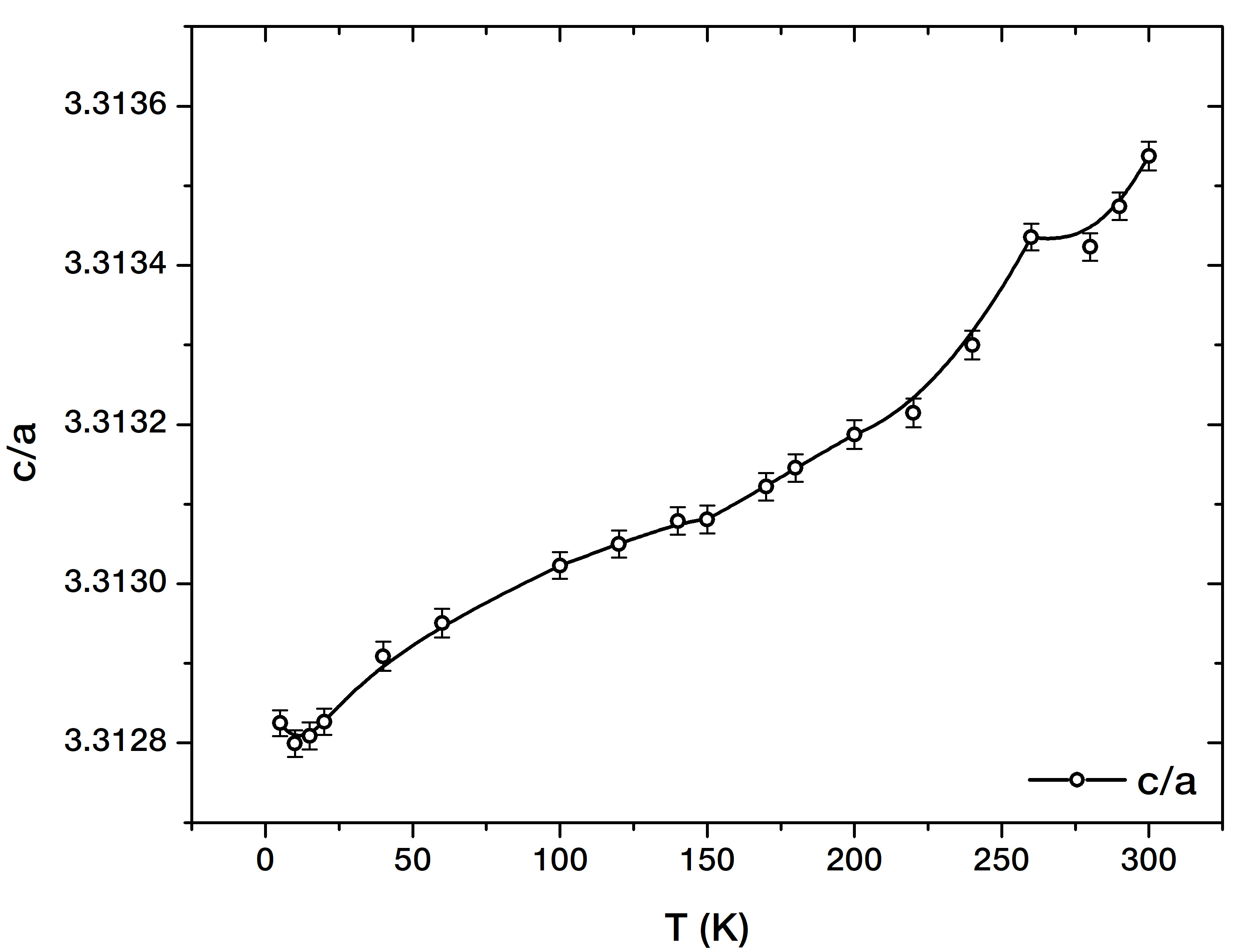}
	\caption{Thermal evolution of the axial ratio $c/a$ obtained from $T-$dependendent SXRD. The antiferromagnetic transition is again seen as the feature at 250 K, along with subtler changes at lower temperatures $\sim$150 K and 10 K.}
	\label{Tdepuc2}
\end{figure}

Fig.\ref{MHNa} shows the $M(H)$ isotherms measured at various temperatures between 2 - 400 K. Overall, an antiferromagnetic behaviour is observed with a linear relationship between $M$ and $H$ without any saturation upto 8 T. A small loop opens up at low temperatures suggestive of weak ferromagnetic correlations. This is in line with the subtle upturn observed in $\chi_{DC}$ below 12.5 K. The origin of both the rise in $\chi_{DC}$ and the loop opening at 2 K could be a weak ferromagnetic contribution contributing to the magnetic response at low temperature. Insets of (b) and (c) show a sharp change in the slope $dM/dH$ at about $\pm$8.5 kOe for $T=$ 2 K and 35 K. This feature is suppressed as $T$ increases and for $T=$ 125 K, 200 K and 225K, subtle step-like features are visible, reminiscent of metamagnetic behaviour. This also manifests in the d$M$/d$H$ in the low-field limit (Fig.\ref{MHNa}(l)) where distinct features can be seen at $\pm$ 100 Oe which are washed out both at higher and lower temperatures. 

\begin{figure}[ht]
	\centering
	\includegraphics[width=0.8\columnwidth, trim=2cm 0 0 0]{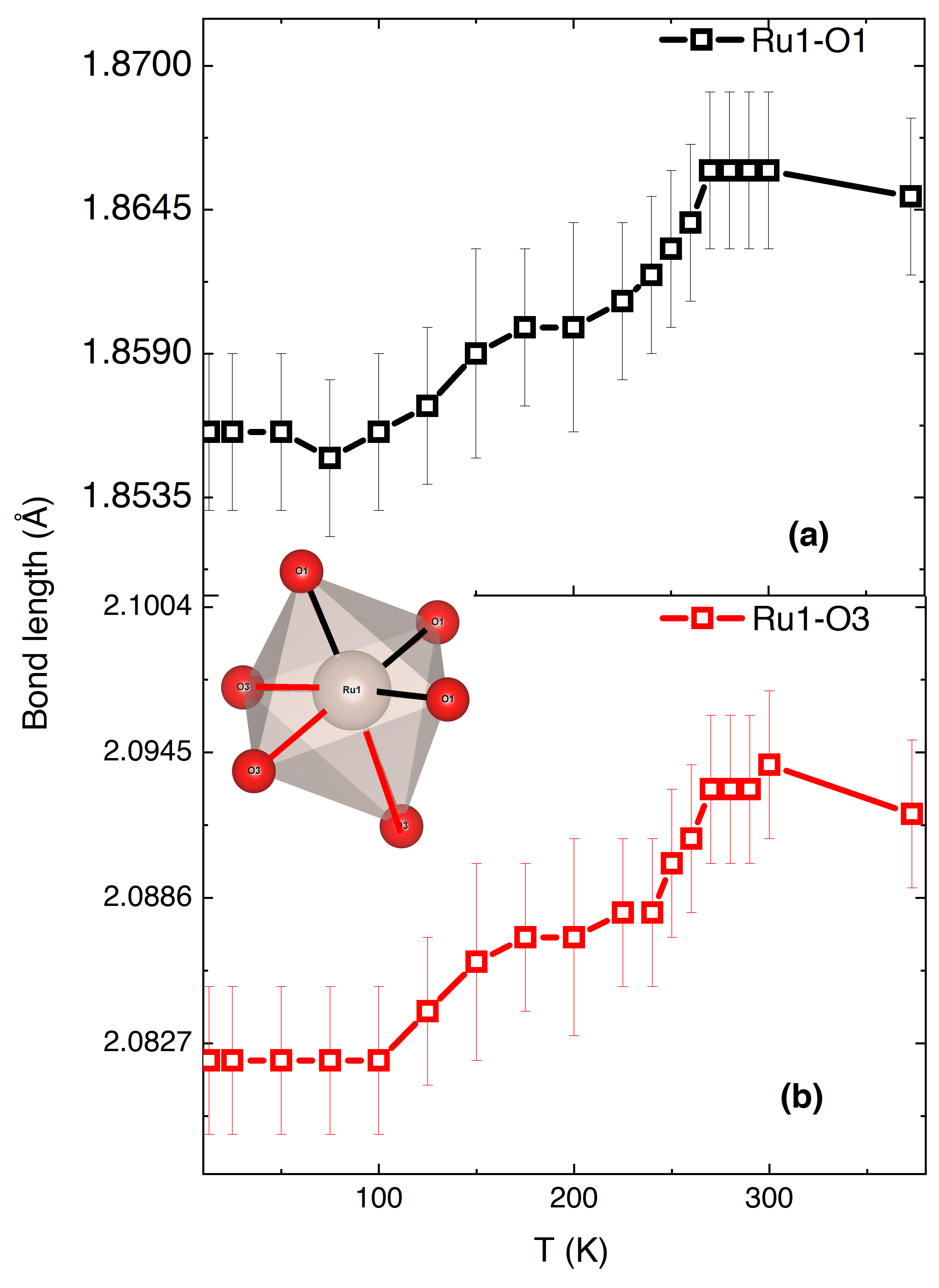}
	\caption{Thermal variation of the Ru1---O octahedral bond lengths, highlighting the AFM transition at $\sim$ 250 K. The colours of the curves correspond to the colour coding on the bonds shown in the pictorially represented octahedra.}
	\label{Tdepbonds}
\end{figure}

\begin{figure}[ht]
	\centering
	\includegraphics[width=0.8\columnwidth, trim=1.5cm 1cm 0.5cm 0]{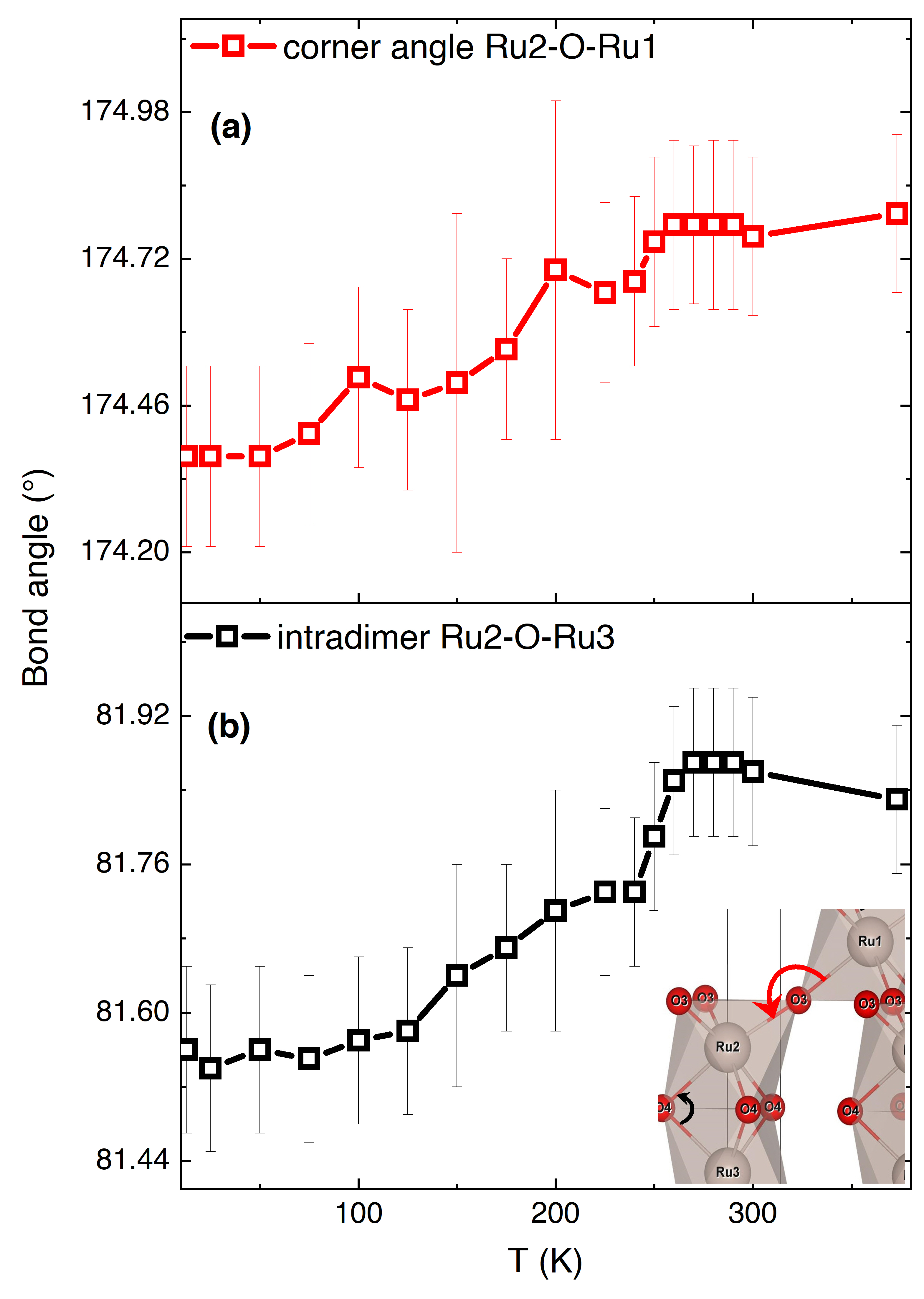}
	\caption{$T$ dependence of the (a) inter- and (b) intra-dimer Ru---O---Ru bond angles, also pictorially represented and colour coded to match the colour of the plotted curves. A marked change across 250 K can be seen in each plot.}
	\label{Tdepbonds2}
\end{figure}

\begin{figure}[ht]
	\centering
	\includegraphics[width=0.8\columnwidth]{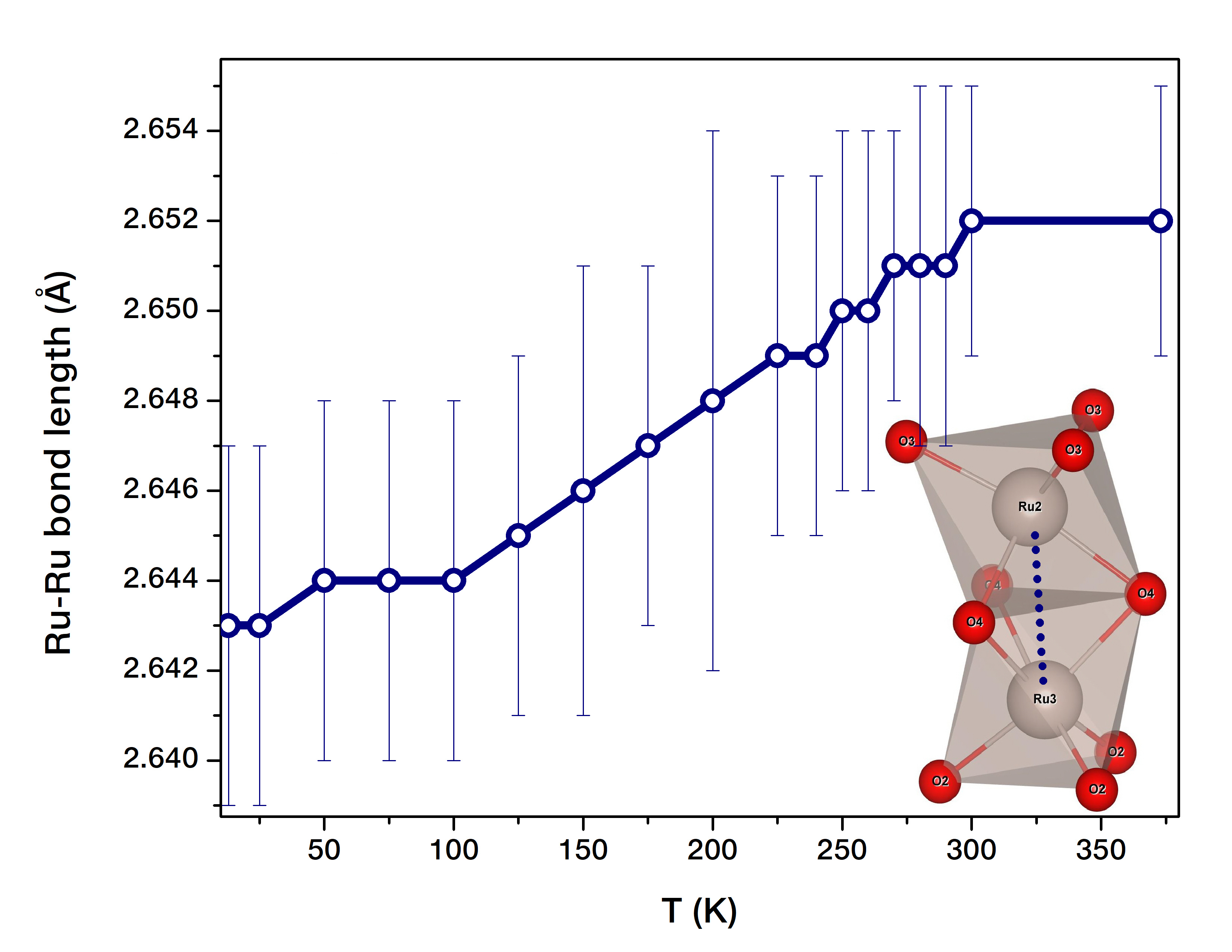}
	\caption{$T$ dependence of the Ru-Ru bond length (visualized in the dimer unit drawn on the plot). The distance between the two Ru ions within the dimer only decreases on cooling and no distinct features corresponding to the AFM transition are observed.}
	\label{ruru}
\end{figure}

Tracking the variation in the coercivity ($H_c$) and remnant magnetization ($M_R$) with temperature brings out the sharp change in both parameters across $T_N$ and subsequent increase below $\sim$ 100 K. This appearance of symmetric dips in d$M$/d$H$ at low fields in an intermediate range of temperatures within the antiferromagnetically ordered phase suggests the possibility of a field-induced domain reorientation or a spin-flop like phenomenon. It will be interesting to perform further magnetic field-dependent measurements on this system to closely study its metamagnetic behaviour.

The analysis of structural parameters at various temperatures through synchrotron x-ray and neutron powder shed light on the variation in bond parameters that lay the foundation of the exchange pathways within this system. The strong spin-lattice coupling is revealed in Figs.\ref{Tdepuc} and \ref{Tdepuc2} where the lattice parameters and the axial ratio ($c/a$) both show an anomaly across T$_N$. As shown in Fig.\ref{structure}(e), there are several exchange pathways at play here - the intradimer direct exchange between Ru2 and Ru3, the intradimer superexchange Ru2---O4---Ru3, the Ru1---O3---Ru2 superexchange, and the Ru1---O1---O2---Ru3 super-superexchange. 

The $T$-dependence of Ru1--O octahedral bonds is shown in Fig.\ref{Tdepbonds}. The corner shared Ru octahedra is significantly distorted as visible in the difference between the Ru1---O1 and Ru1---O3 bond lengths and shows marked changes across the 250 K in agreement with the magnetic measurements. Both bonds shrink between 270 and 225 K, plateau between 225 and 175 K, continue to further shrink below 175 K till $\sim$ 100 K below which no $T-$ dependence is observed. The superexchange bond angle within the dimer (Ru2---O4---Ru3) and that between the corner sharing Ru ions (Ru1---O3---Ru2) are shown in Fig.\ref{Tdepbonds2} (a) and (b), respectively. The AFM transition is clearly highlighted in the intradimer angle (Fig.\ref{Tdepbonds2} (b)) which closely follows the Ru1---O3 bond angle (Fig.\ref{Tdepbonds} (b)). These features can be correlated to the subtle changes in the rate of decline of $\chi_{mol}$ with $T$ as seen in Fig.\ref{MTNa}. The Ru-Ru overlap strengthens on cooling as evidenced by a monotonic decrease in the Ru2---Ru3 bond length (Fig.\ref{ruru}). This bond seems to remain unaffected by the antiferromagnetic transition and no corresponding features are observed in the thermal variation of this bond length within error limits.

\begin{figure}[h]
	\centering
	\includegraphics[width=\columnwidth]{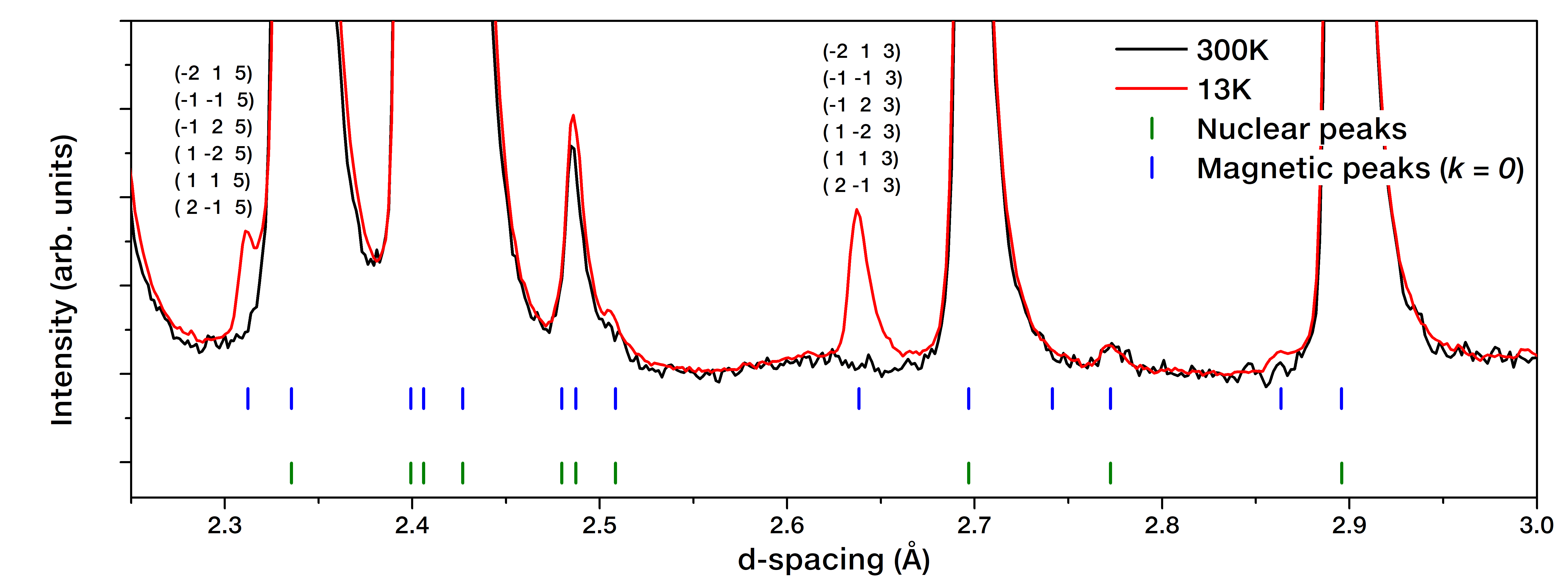}
	\caption{NPD data collected at $T=13$ K superimposed on room temperature data - highlighting the additional reflections - corresponding to the antiferromagnetic structure. The \emph{hkl} indices of the planes contributing to these intensities are listed above the peaks.}
	\label{magpks}
\end{figure}


The appearance of new reflections not allowed by the crystal symmetry $P6_3mc$ confirmed the existence of long-range magnetic order in the system. Fig.\ref{magpks} shows comparison between the diffraction patterns collected at 300 and 13 K, clearly highlighting the two additional reflections at $d=2.637$ \AA{} and $2.311$ \AA{} which are indexed by the k-vector \textit{\textbf{k}} = (0,0,0). These are the two most prominent peaks and no other reflections are observed at higher d-spacings. 

In order to solve the magnetic structure corresponding to these reflections, we performed magnetic representational analysis using the program {\footnotesize SARA\textit{h}} \cite{wills2000}. Although the three Ru sites are symmetrically unique, their site multiplicities are same (each generates two equivalent positions) which means that their magnetic representations ($\Gamma_{Mag}$) are the same. For each site, we obtained $\Gamma_{Mag} = 0\Gamma_1^1 + 1\Gamma_2^1 + 0\Gamma_3^1 + 1\Gamma_4^1 + 1\Gamma_5^2 + 1\Gamma_6^2$. The point symmetry and Shubnikov Group corresponding to each basis vector for each surviving irreducible representation (IR) is given in Table \ref{bv_symmetry_vector_table_3}.

\begin{table}[H]
	\small
	\centering
	\begin{tabular*}{\columnwidth}{@{\extracolsep{\fill}}cccll}
		\hline
		IR  &  BV  &  Point Group & \multicolumn{2}{c}{Shubnikov Group} \\
		\hline
		$\Gamma_{2}$ & $\psi_{1}$ &      6 & P6$_3$m'c' & $(P6_3m'c'  , BNS \# 186.207)$  \\
		$\Gamma_{4}$ & $\psi_{2}$ &      6 & P6$_3$'m'c & $(P6_3'm'c  , BNS \#  186.205)$  \\
		$\Gamma_{5}$ & $\psi_{3}$ &      6 & C/B & \\
		& $\psi_{4}$ &     3m & P1 & $(P1  ,  BNS \# 1.1)$  \\
		$\Gamma_{6}$ & $\psi_{5}$ &      6 & C/B & \\
		& $\psi_{6}$ &     3m & P1 & $(P1  ,  BNS \# 1.1)$  \\
		\hline
	\end{tabular*}
	\caption{The point symmetries and Shubnikov space groups associated with each basis vector calculated from the parent space group $P 6_3 m c$ with $\mathbf{k} =(0,~ 0,~ 0)$. C/B indicates that a change of basis is needed to define the Shubnikov group, i.e. the axis definitions no longer match those of the parent structure used for these calculations. For example, a rhombohedral cell in the hexagonal setting with 3 centring translations being changed to a C-centred monoclinic cell with only the C-centring.}
	\label{bv_symmetry_vector_table_3}
\end{table}

\begin{figure}[h]
	\centering
	\includegraphics[width=\columnwidth]{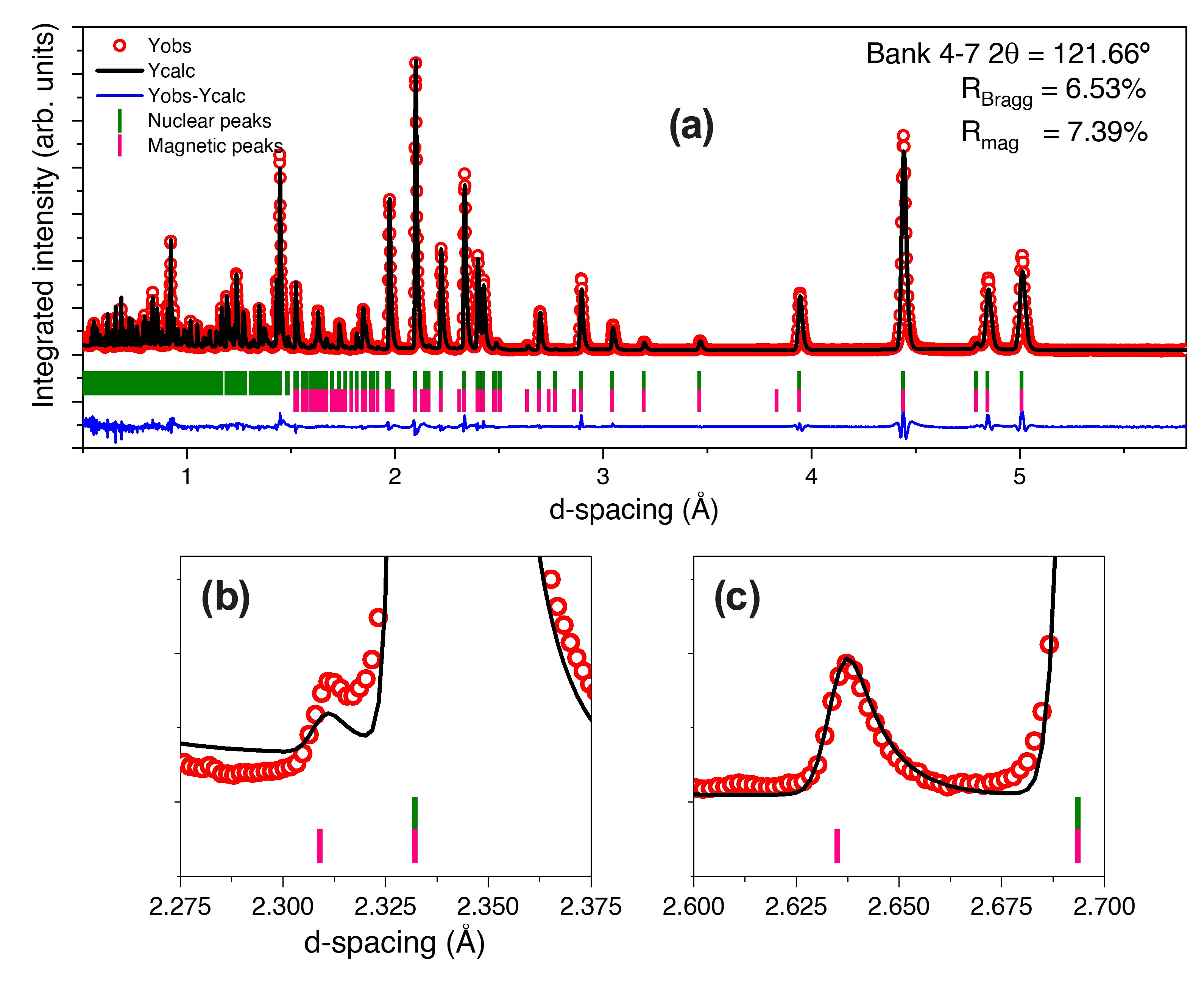}
	\caption{(a) Rietveld fit of the IR $\Gamma_4$ magnetic model (magnetic space group $P6_3'm'c$) to neutron powder diffraction data collected at 13 K. (b), (c) Zoomed plots showing agreement of the model to the magnetic reflections.}
	\label{13Kfit}
\end{figure} 

The best fit to the magnetic peaks is obtained using the IR $\Gamma_{4}$ corresponding to the magnetic space group $P6_3'm'c$ (No. 186.205), implying that the time-reversal symmetry is broken along the $c$ and the $a$ axes due to magnetic ordering. The Rietveld fit of this model to the observed diffraction pattern at 13 K is shown in Fig.\ref{13Kfit}(a). The accuracy of the model is confirmed by the reasonable magnetic R-factor (R$_{mag}$) of 6.33\% and the fits to the magnetic reflections (Fig.\ref{13Kfit}(b) and (c)). It is worth noting that the choice of IR was made based on the best fit obtained with freely varying Ru moments. This unconstrained model had a slightly better R$_{mag}$ ($\sim$5.7\%) but produced a structure assigning nearly identical moment sizes on Ru1 and Ru2 sites while leaving Ru3 with just about half that value. This is physically unlikely since Ru2 and Ru3 are much closer to each other and are likely to have a strong AFM correlation. Constraining all Ru sites to have the same moment values resulted in a structure with ferromagnetic alignment within the dimers (R$_{mag}$$\sim$12.3\%), inconsistent with magnetic data. Constraining Ru2 and Ru3 moments to have the same magnitude, keeping the same orientations as obtained from the unconstrained model, resulted in identical visual fits, with only a minor increase in the R$_{mag}$ (5.7 $\rightarrow$ 6.3 \%) indicative of a physically sensible model. In contrast, R$_{mag}$ for $\Gamma_{2}$ and $\Gamma_{6}$ R$_{mag}$ were found to be 13.5\% and 117\%, respectively; while $\Gamma_{5}$ led to divergence of the fit.

\begin{figure}[h]
	\centering
	\includegraphics[width=\columnwidth]{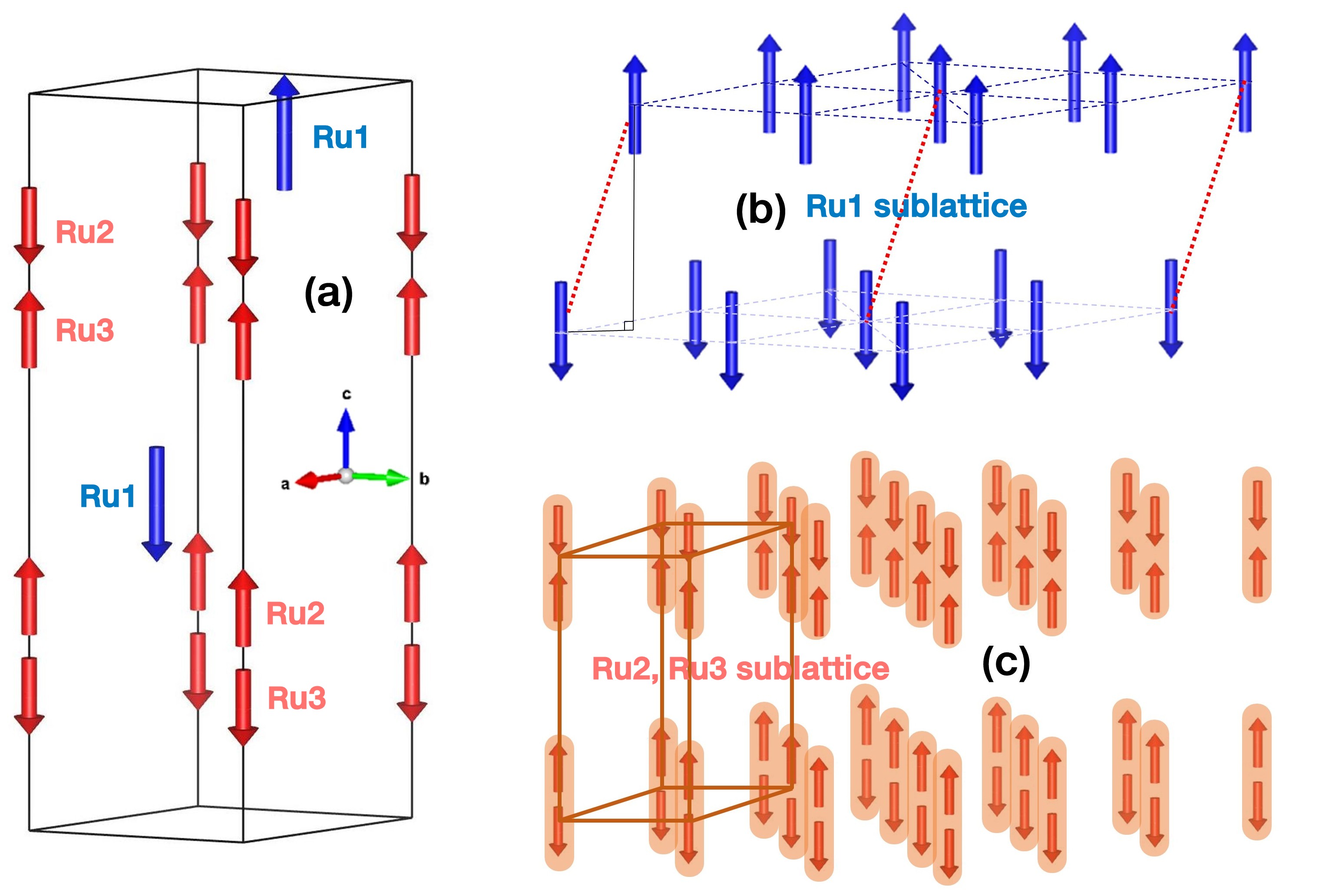}
	\caption{(a) The A-type AFM structure of Ba$_3$NaRu$_3$O$_{12}$ at 13 K. (b) The staggered arrangement of the Ru1 triangular layers, AFM coupled to each other along $c$. (c) The cubic packing of the Ru2,3 dimers - characterized by intra-dimer and inter-layed AFM coupling. Both sublattices are intra-layer FM coupled.}
	\label{magstrucNa}
\end{figure}

\begin{figure}[h]
	\centering
	\includegraphics[width=\columnwidth, trim = 0 0 0 1cm]{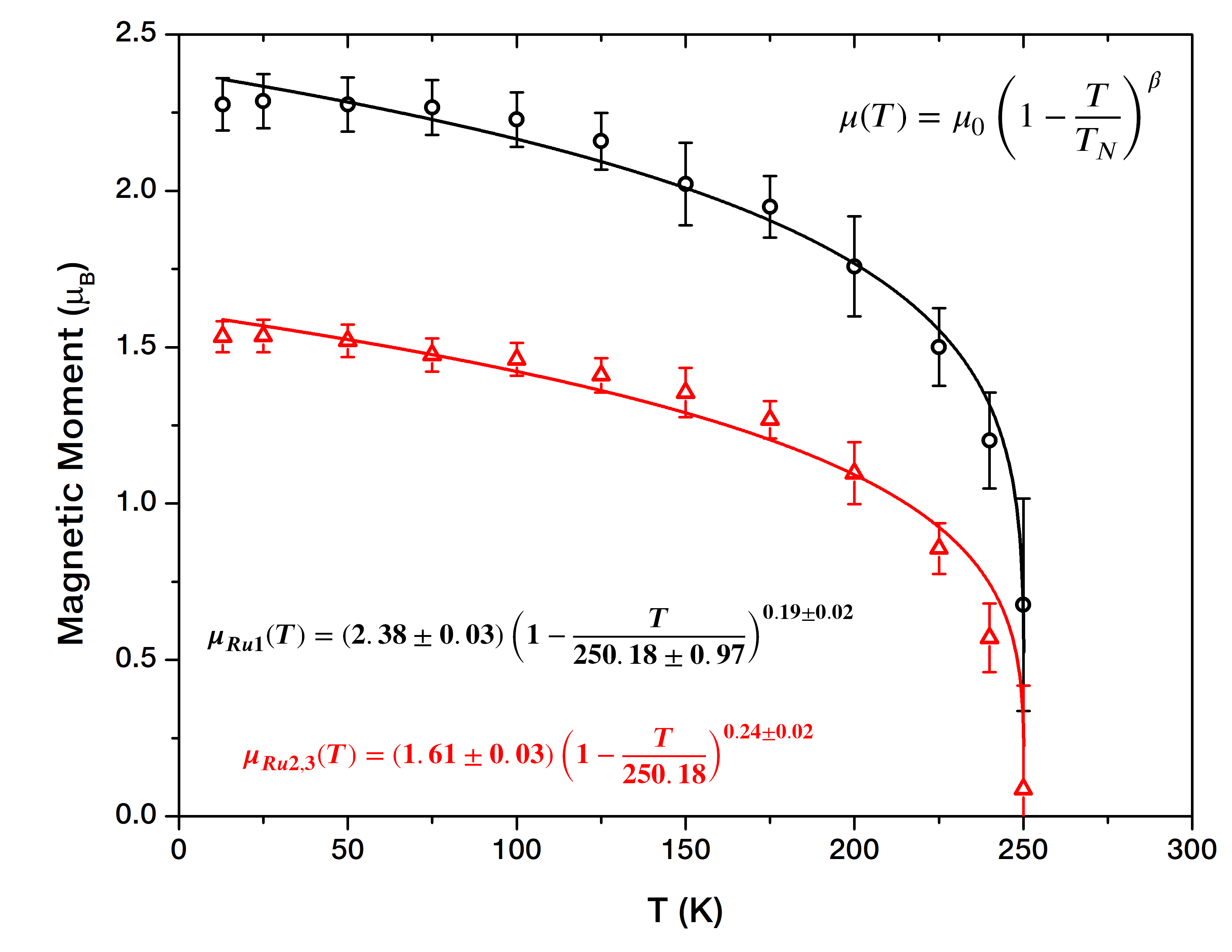}
	\caption{Growth in Ru magnetic moment on cooling. Moment values at each $T$ obtained from fitting $T-$dependent NPD data to the magnetic space group $P6_3'm'c$. $\mu (T)$ curves fitted to mean field equation (shown on plot) to obtain the critical exponent $\beta$ and $\mu (T=0)$.}
	\label{moments}
\end{figure}

The magnetic structure (Fig.\ref{magstrucNa}(a)) exhibits an A-type arrangement of spins which are ferromagnetically coupled within each layer, while AFM coupling is present between the NN layers. All the spins are aligned collinear to the crystalline easy axis $c$, a testament to the strong influence of the structure on the magnetic order. An interesting aspect of this structure is shown in Fig.\ref{magstrucNa}(b) and (c), which is the difference in the packing of the different layers. If the Ru1 and Ru2,3 sublattices are separated out, one can observe that the Ru1 layers have a staggered arrangement owing to the $6_3$ axis running along $c$, while the Ru2,3 have a cubic packing. Similar A-type orderings with spins locked in the $ab$-plane, have been found in other layered perovskite systems with cubic packing like Ba$_3$Fe$_2$(W,Te)O$_9$\cite{tang2017magnetic,ivanov2004nuclear} and NdBaMn$_2$O$_6$\cite{akahoshi2004charge}. Ba$_2$CoTeO$_6$ also shows an A-type order but with spins canted at an angle of 24.5\textdegree with respect to the $c$-axis\cite{ivanov2010neutron}. Ba$_3$NdRu$_2$O$_9$ shows a FM order of Nd spins collinear with the $c$-axis below 24 K which cant below 18 K accompanied by an AFM order of the dimer Ru spins\cite{senn2013spin}. This exemplifies the vast variation in magnetic stuctures possible and its strong dependence on the underlying crystal structure and the type of TM ions making up the exchange network. We note that this work is the first conclusive solution of the ground state magnetic structure of of a B-site substituted non-trimer quadruple perovskite.

\begin{figure}[h]
	\centering
	\includegraphics[width=\columnwidth]{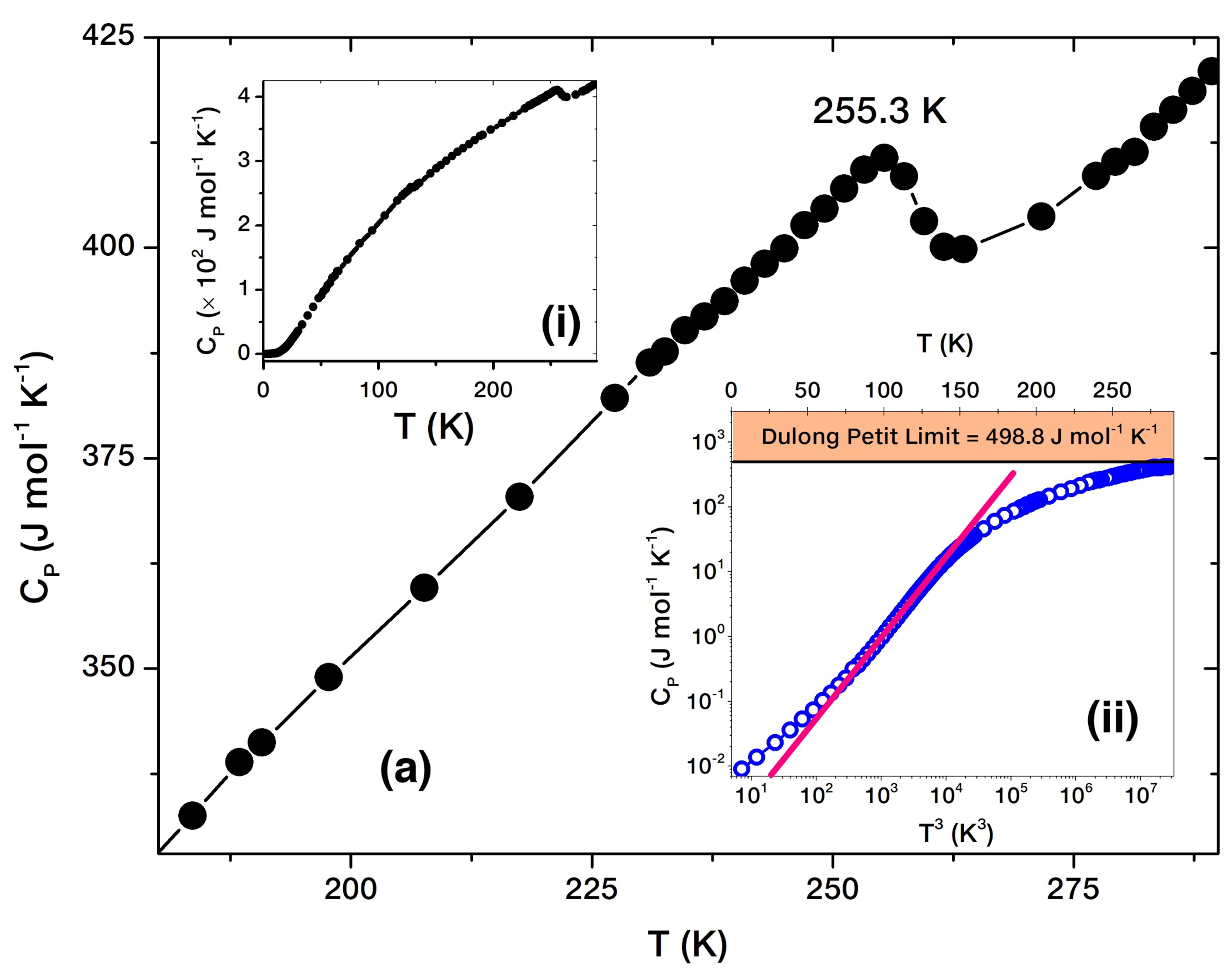}\\
	\includegraphics[width=\columnwidth]{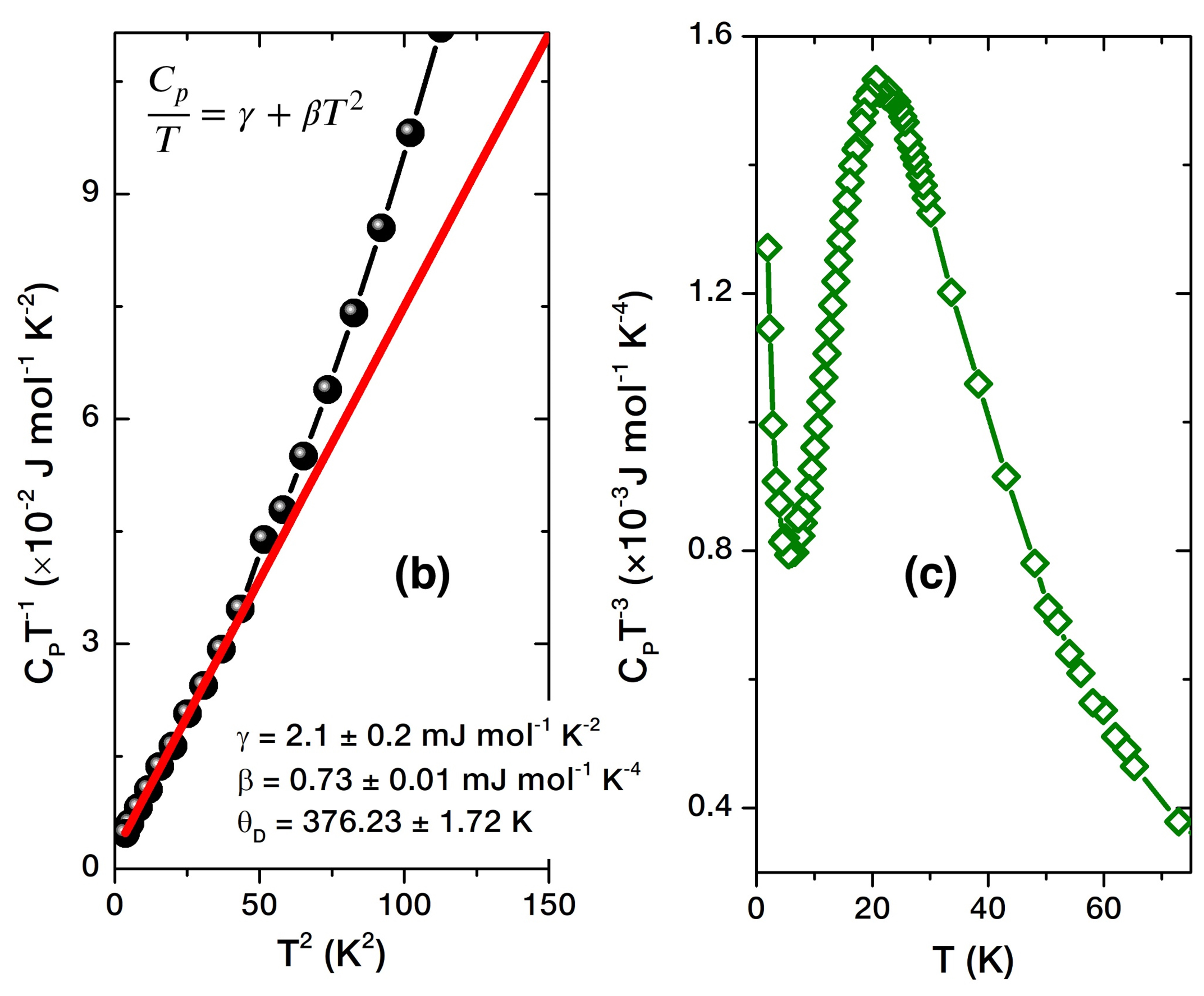}
	\caption{(a) Temperature dependence of specific heat capacity of Ba$_4$NaRu$_3$O$_{12}$ in the vicinity of T$_N$. Inset (i) shows $C_P(T)$ for the whole $T$ range (2 - 300 K) and inset (ii) shows a plot of $C_P$ vs $T^3$ showing the range of temperatures for which the Debye term is dominant. (b) A plot of $C_P/T$ vs $T^2$ with the linear fit along with the parameters obtained from the fit. (c) A graph showing the variation in $C_P/T^3$ as a function of T.}
	\label{Cp}
\end{figure}

Fitting the model to NPD patterns collected at different temperatures allows us to track the development of the magnetic phase as a function of temperature. The rise in moment values at each magnetic site with cooling, is shown in Fig.\ref{moments}. Both Ru moments record a qualitatively similar curve with Ru1 saturating at $\sim$ 2.25 $\mu_B$ and Ru2 (= Ru3) at $\sim$ 1.5 $\mu_B$. We fit the growth of moment below $T_N$ to a power-law function to ascertain the magnetic order parameter exponent $\beta$ and the temperature of the phase transition. The mean-field equation is:\[\mu(T) = \mu_0 \bigg(1-\frac{T}{T_N}\bigg)^{\beta}\]
where $\mu_0$ is the magnetic moment at $T = 0$, $T_N$ the temperature of the phase transition and $\beta$ the critical exponent capturing the behaviour of the order parameter ($\mu$) at criticality. For Ru1, the parameters obtained are: $\mu_0 = $2.38$\pm$0.03 and for Ru2 (=Ru3), $\mu_0 = $1.61$\pm$0.03, elucidating the effect of Ru-Ru direct interaction within the Ru$_2$O$_9$ facilitating the suppression of the individual Ru values. The critical exponent for Ru1 sites is $\beta = $ 0.19 $\pm$ 0.02, and that for Ru2,3 sites is $\beta = $ 0.24 $\pm$ 0.02 (using T$_N$=250.18 K from the $\mu_{Ru1}(T)$ fit) placing Ba$_4$NaRu$_3$O$_{12}$ well within the universality window of two-dimensional critical exponents \cite{taroni2008universal}, with its lower limit marked by the Ising 2D model ($\beta = 1/8 = 0.125$). Values of $\beta$ for 3D magnetic models is expected to be much larger ($\beta = $0.33), while Fe(NCS)$_2$(pyrazine)$_2$, a $S=1/2$ 2D molecular antiferromagnet is reported to have the exact same value of $\beta$ \cite{bordallo2004}. Similar values of $\beta=$0.18$\pm$0.03 and $\beta=$ 0.20 $\pm$ 0.02 are reported for PdCrO$_2$, an $S=3/2$ triangular lattice antiferromagnet\cite{takatsu2009critical}, and La$_2$CoO$_4$, a 2D $S=1/2$ layered material with longitudinal Ising spin correlations\cite{yamada1989}. This attests to the layered, quasi-2D nature of the magnetism in Ba$_4$NaRu$_3$O$_{12}$ with ferromagnetic triangular layers of Ru ions are antiferromagnetically linked along the easy-axis ($c$) with out-of-plane collinear spins.

The specific heat measured as a function of temperature (C$_P$(T)) is shown in Fig.\ref{Cp}(a). A clear $\lambda$-type anomaly is visible at 255 K, confirming the antiferromagnetic transition at $T_N$. The small magnitude of the peak likely suggests that the entropy change associated with this transition is small. This observation also matches up with the absence of any symmetry lowering in the crystal structure of the compound as a function of temperature. Inset (i) of Fig.\ref{Cp} shows the data in the full $T$ range highlighting the absence of any more features at lower temperature. This probably means that the octahedral distortions only affect the magnetic ground states and the overall energy of the state remains impervious to those local changes. Inset (ii) of Fig.\ref{Cp} highlights the region where Debye behaviour is observed, and a linearity is captured in the $C_P$ vs $T^{\,3}$ plot. At high temperature $C_P$ can be seen to approach the Dulong-Petit Limit, as expected. 

Fig.\ref{Cp}(b) shows the low-temperature linear fit to $C_P/T$ vs $T^2$, to estimate the electronic and lattice contributions to the specific heat. We obtain a Debye Temperature $\theta_D =$ 376.23 $\pm$ 1.72 K using \[ \theta_D = \bigg( \frac{12 \pi^4 p R }{5 \beta} \bigg)^{1/3} \] consistent with reports for other layered perovskites \cite{igarashi2015,radtke2013}. Here, $p$ is the number of atoms per formula unit (20 for quadruple perovskites) and $R$ the gas constant. The values obtained for $\gamma = $ 2.1 $\pm$ 0.2 mJ mol$^{-1}$K$^{-2}$ and $\beta = $ 0.73 $\pm$ 0.01 mJ mol$^{-1}$K$^{-4}$ are both minuscule, implying that most of the entropy of the system is lost in approaching low temperatures. This is further corroborated in the plot of $C_P/T^3$ vs $T$ where a peak in $C_P/T^3$ is observed at $\sim$ 21 K. This excess phonon specific heat is a common feature in many perovskite systems and is attributed to the presence of low-lying optical (Einstein) modes that cause a departure from the phononic density of states and promote a pure $T^3$ dependence of $C_P$ at $T<\theta_D$. The position of the peak is said to be a manifestation of the \emph{boson peak} \cite{dangelo2009} and can be used to estimate the vibration energy of the largest contribution to $C_P$ at the peak temperature under the dominant phonon approximation \cite{berman1976thermal}. For the observed peak in $C_P/T^3$ at $\sim$20.8 K, we obtain a vibrational energy of $\sim$ 55 cm$^{-1}$ (using \(\hbar \omega = 3.8 k_B T_{max}\))\cite{berman1976thermal}.  

\begin{figure}[h]
	\centering
	\includegraphics[width=\columnwidth]{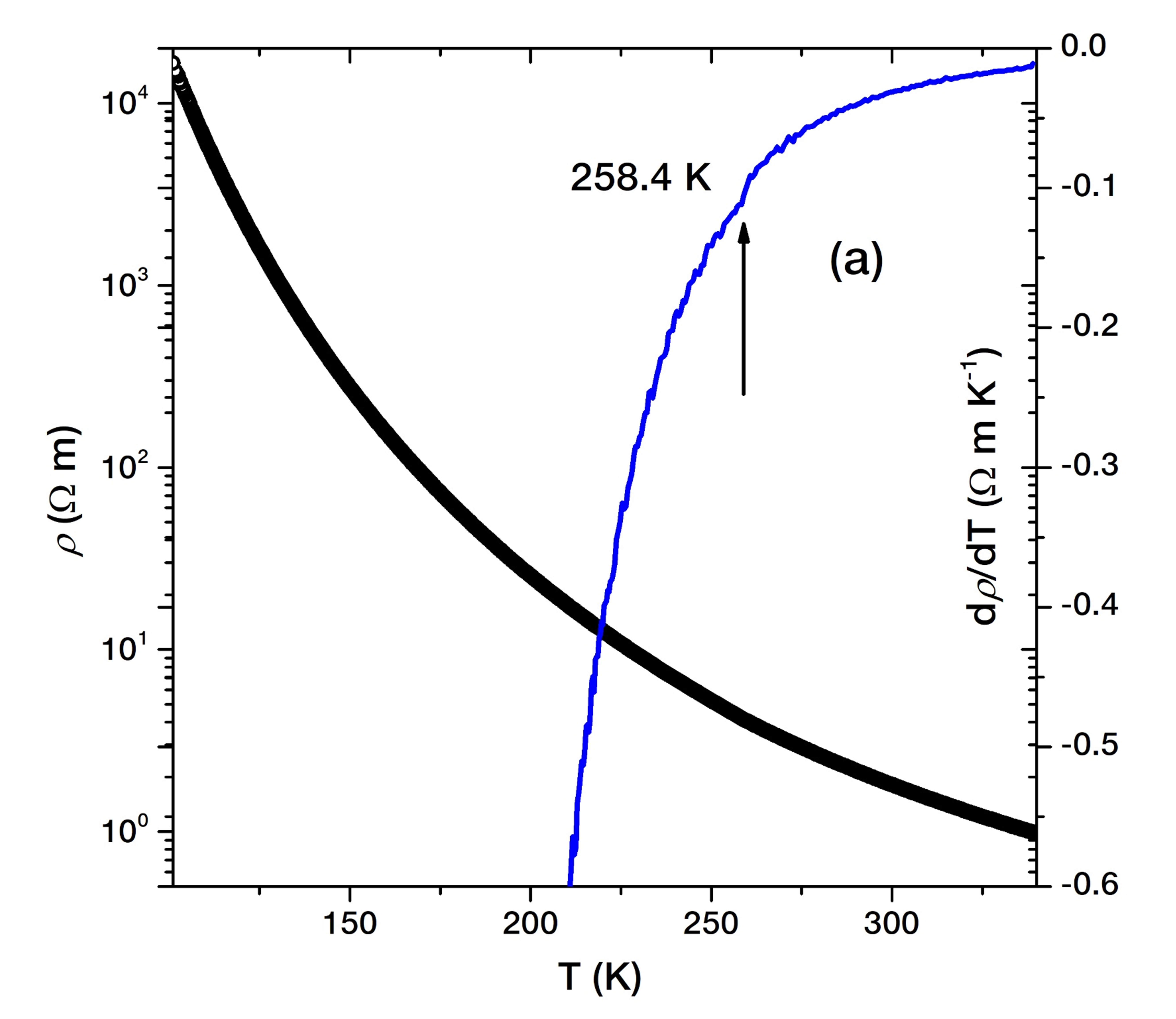}
	\includegraphics[width=\columnwidth]{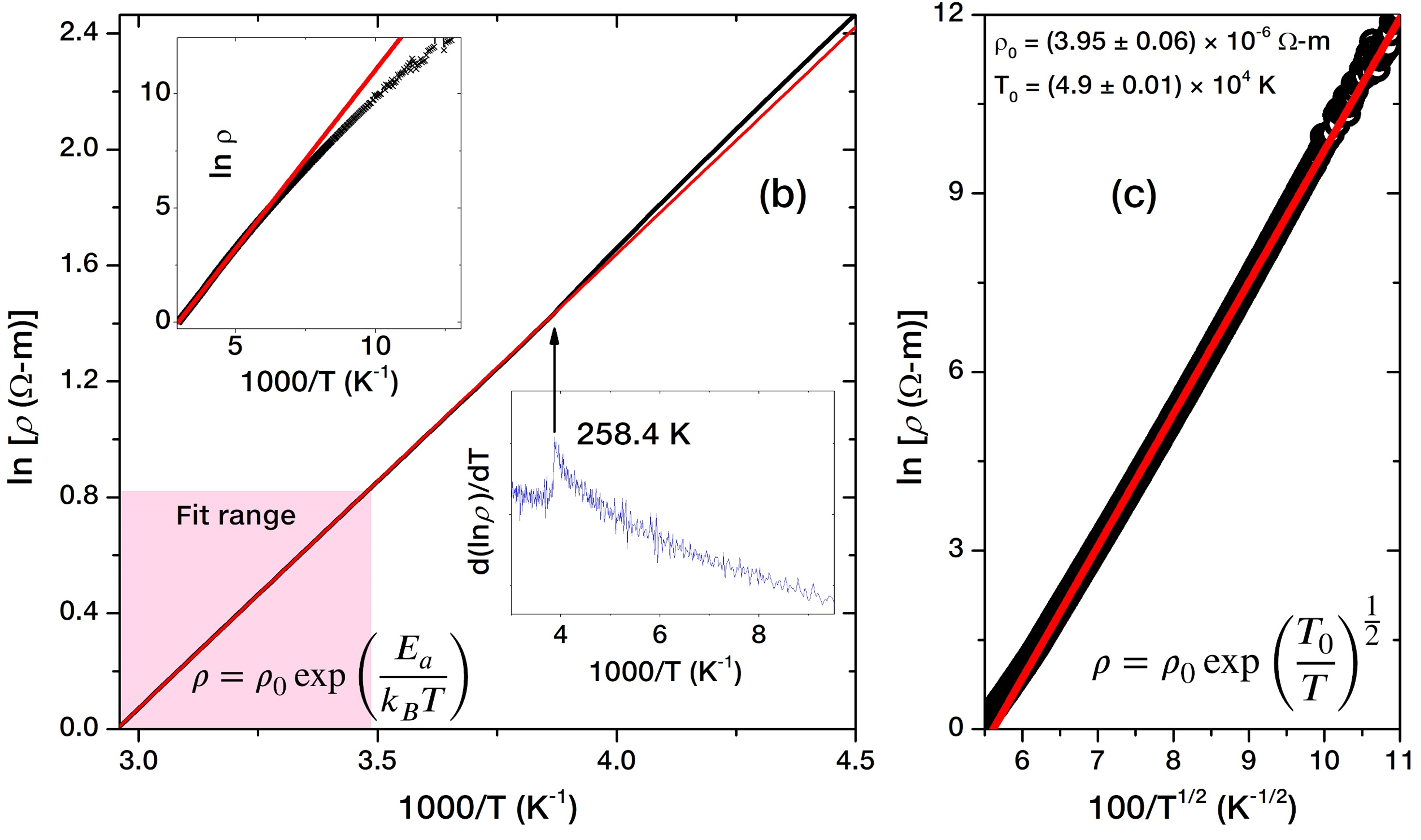}
	\caption{Transport behaviour of Ba$_4$NaRu$_3$O$_{12}$.(a) $\rho$ vs $T$, showing a semiconducting behaviour, and the derivative d$\rho$/d$T$ shows the discontinuity at $T_N$. (b) Linear fit to ln $\rho$ vs $10^3/T$ for $T>T_N$. Insets show the deviation from linearity and the sharp change in the slope below $T_N$. (c) Linear fit to the Efros-Shklovskii Variable Range Hopping (ES-VRH) model below $T_N$.}
	\label{RTNa1}
\end{figure} 

Fig.\ref{RTNa1}(a) shows the semiconducting nature of the material along with a small discontinuity observed in the derivative $d\rho/dT$ at $T_N$. The charge transport shows a thermally activated behaviour above $T_N$ with a linear ln$\rho$ against $T^{-1}$. Fig.\ref{RTNa1}(b) shows the linear fit in the $T$ range 333 - 287 K. Deviation from this linearity begins close to the magnetic transition and is depicted more clearly in the zoomed inset of Fig.\ref{RTNa1}(b) and the derivative plot $d(ln\rho)/dT$ vs $10^3 T^{-1}$ marking the onset of a declining slope below $T_N$. The linear fit returned an activation energy of $E_a$ = 135.08 $\pm$ 0.04 meV and the minimum resistivity $\rho_0$ = 9.75 $\pm$ 0.01 m$\Omega$-m.  

\begin{figure}[h]
	\centering
	\includegraphics[width=\columnwidth]{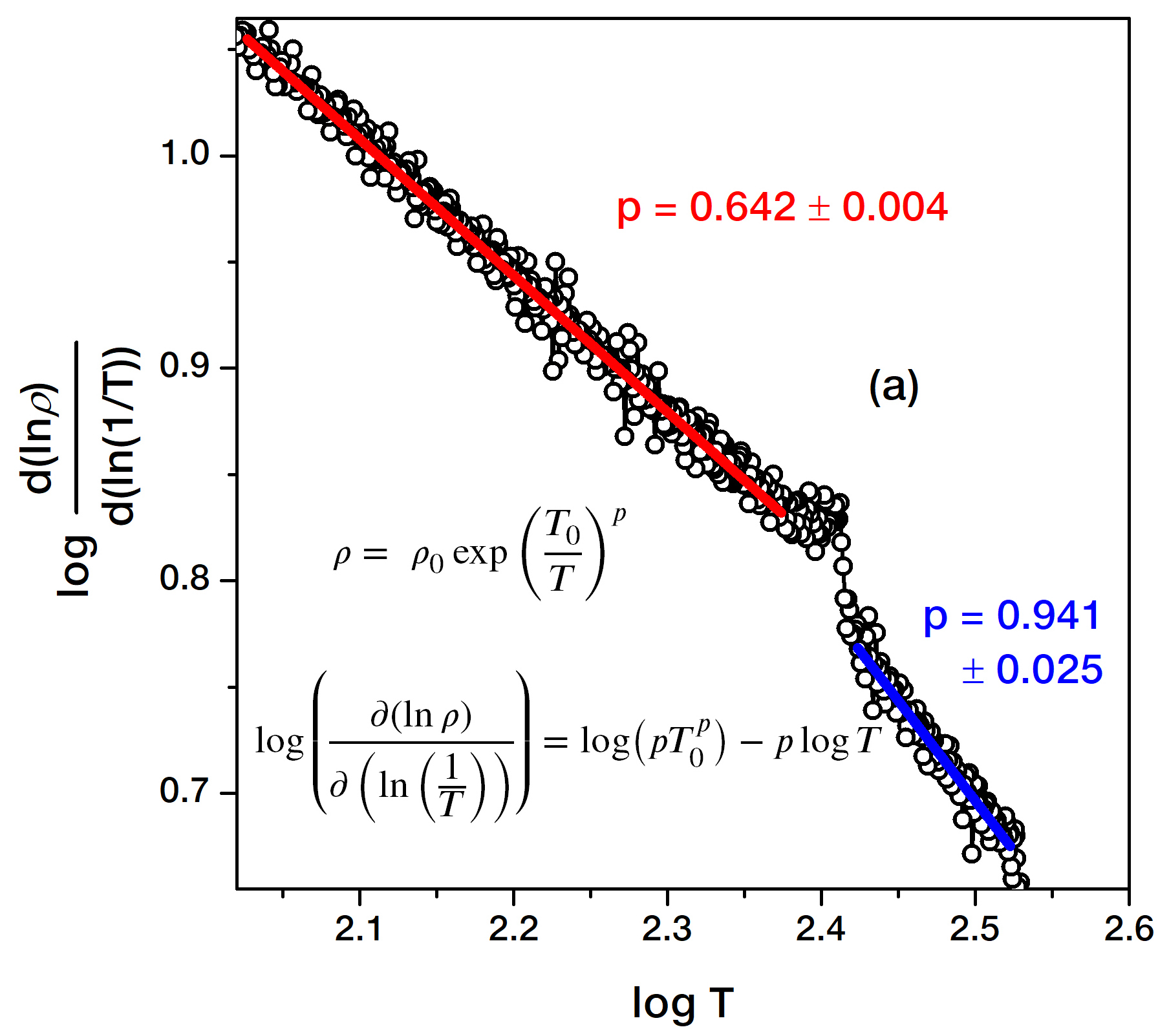}
	\includegraphics[width=0.9\columnwidth]{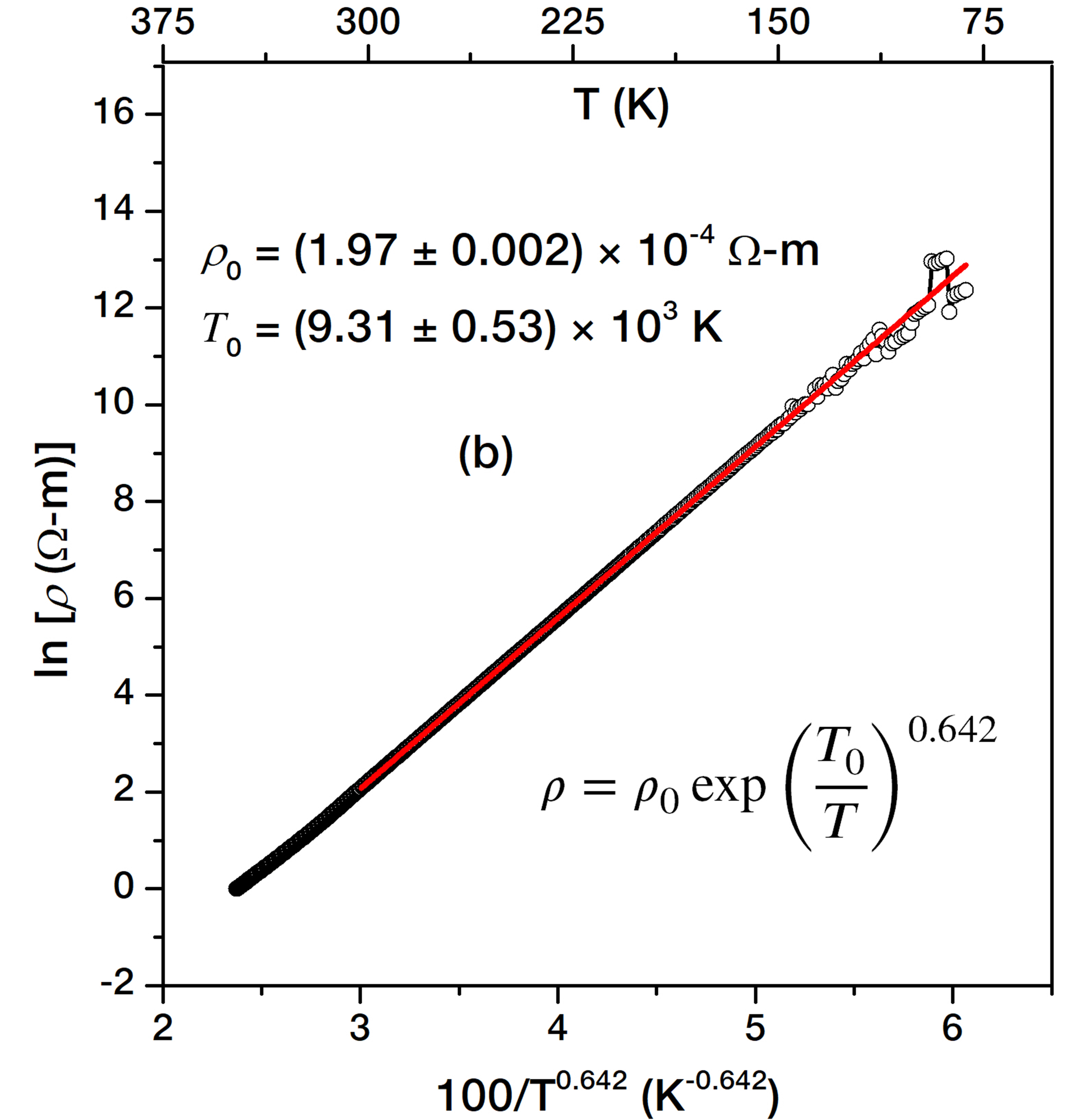}
	\caption{(a) log-log plot of $d(ln\rho)/d(ln(1/T))$ showing the two distinct transport regimes characterized by different slopes ($p$). (b) ln$\rho$ vs 100/$T^{0.642}$ yielding an improved linear fit with a lower $T_0$}
	\label{RTNa2}
\end{figure}

Below $T_N$, when ln$\rho$ was found to be non-linear with respect to Mott's Variable Range Hopping  (VRH) behaviour \cite{Mott1969}, we employed the generalized VRH law given by:
\begin{equation}
	\rho = \rho_0 exp\bigg(\frac{T_0}{T}\bigg)^p
\end{equation}
where $\rho_0$ and $T_0$ define the minimum resistivity at T$\rightarrow$$\infty$, and characteristic temperature of the hopping behaviour, respectively. $p$ is the \textit{hopping exponent}, which takes the value 1/3 and 1/4 in Mott's law  \cite{Mott1969} for 2D and 3D systems, respectively, having noninteracting electrons and constant density of states $N(\epsilon) = N_0$. However, in real materials, electron-electron interactions are often important and their inclusion modifies the Mott-VRH to the Efros-Shklovskii (ES) VRH \cite{efros1975} with $p = 1/2$ for both 2D and 3D systems. The density of states in this model has a quadratic dependence on the energy close to the Fermi energy ($\epsilon_F$) given by\cite{shklovskii1984, shklovskii2024}:
\begin{equation}
	N(\epsilon) = \frac{\pi}{3}\frac{\kappa^3(\epsilon-\epsilon_F)^m}{e^6}
\end{equation}
with $m=2$ in a 3D semiconductor where, $\kappa$ is the dielectric constant of the material and encapsulates the effect of Coulomb interaction between the electrons. This leads to the opening of a soft gap in the single particle density of states, the width of which is known as the \emph{Coulomb gap}. Fitting the ln$\rho$ to $1/T^{1/2}$ returns a reasonably linear fit across 7-8 orders of magnitude, however, the value $T_0 = 4.3 \times 10^4$ K obtained is significantly high. Similar results, with even higher values of $T_0$ have been reported for the $5d$ double perovskites A$_2$MnReO$_6$ (A = Ba, Sr, Ca)\cite{fisher2008}.

Now, the hopping exponent can be \emph{a priori} extracted from the $\rho$ vs $T$ curve by plotting log$\frac{d(ln\rho)}{d(ln(1/T))}$ vs log$T$, the slope of which returns $-p$ \cite{massey1995}. As shown in Fig.\ref{RTNa2}(a), the temperature regimes $T>T_N$ and $T<T_N$ are defined by two different slopes. For $T>T_N$, we find $p\approx$1, as discussed earlier, and for $T<T_N$, $p = 0.642 \pm 0.004$, which is higher than the ES value. In a linearized plot (Fig.\ref{RTNa2}(b)), this hopping exponent yields a significantly better fit. Generally, the exponent describing the dependence of $N(\epsilon)$ on $\epsilon$ ($m$) is related to the hopping exponent as: \(p = (m+1)/(m+4)\) \cite{pollak1972}, so that for the ES model, when $p=1/2$, $m=2$ and \(N(\epsilon)\propto(\epsilon-\epsilon_F)^2\). In the present case, the obtained value of $p = 0.642$ yields the relation \(N(\epsilon)\propto(\epsilon-\epsilon_F)^{4.38}\), a far stronger dependence than proposed by ES-VRH. This scenario is likely if multi-particle excitations due to correlation effects are considered instead of the single-particle consideration of the ES model. \cite{pollak1980}. Numerical calculations also have suggested that the Coulomb gap can have a quasiexponential energy dependence in presence of multi-electron transitions \cite{davies1984}. This attests to the strongly insulating low temperature state, where the total energy available to the system is greatly reduced and thus, so are the density of states close to Fermi level, leaving the system with little possibility to conduct. Another interesting aspect here is the lack of existence of the Mott regime. Generally, a gradual crossover from Mott to ESVRH regime is observed on cooling, and the Mott and ES characteristic temperatures $T_0^M$ and $T_0^{ES}$ are related to the Coulomb gap width ($\Delta_{CG}$) as \(\Delta_{CG} = \sqrt{\frac{T_0^3}{T_M}}\)\cite{rosenbaum1991}. But here, we bypass the Mott regime entirely to reach the ES hopping regime. 

\section{Conclusions}

A thorough investigation of Ba$_4$NaRu$_3$O$_{12}$ via structural, magnetic and thermal and electronic transport measurements reveals a robust spin-lattice coupling in its ground state. The system consists of a unique motif uncommon in layered perovskite systems - of three symmetrically unique magnetic ions occupying the centre of a corner octahedra connected to a dimer, isolated by a non-magnetic layer. Despite this separation of magnetic entities, it exhibits a long-range magnetic ordering at a fairly high temperature of $T_N \sim$ 257 K which is corroborated in both specific heat as well as temperature-dependent diffraction measurements, attesting to the strong coupling between the magnetic and lattice degrees of freedom, and the robustness of the exchange mechanism. The magnetically ordered ground state is visualized by the neutron powder diffraction measurements at 13 K and the Ru spins are seen to couple antiferromagnetically along \emph{c} axis and ferromagnetically within the \emph{ab} plane. Thermal dependence of structural parameters indicates the presence of substantial spin-lattice coupling. The magnetization isotherms show an interesting metamagnetic feature in the intermediate $T$-range of 125 - 225 K, concurring with the region where the Ru1---O$_6$ show the maximum change - pointing to some type of spin reorientation. This places Ba$_4$NaRu$_3$O$_{12}$ as an interesting material candidate for a theoretical investigation of the competing interactions influencing its ground state and the reason behind the strength of the antiferromagnetic exchange interaction. The low-$T$ specific heat response is dominated by Einstein (optical) mode vibrations attested to by the peak seen in $C_P/T^{~3}$ plots and the diminished electronic and Debye terms ($\gamma$ and $\beta$). The charge transport shows a transition from a thermally activated behaviour above $T_N$ to a $exp(T^{-p})$ VRH transport with $p = 0.642$, implying the presence of a \emph{soft} Coulomb gap with a stronger energy-dependence than the quadratic dependence suggested by the Efros-Shklovskii model with $p=0.5$. This suggests that multiparticle effects might be active in this system due to strong correlation effects. \\

\begin{acknowledgments}

Ajit Patra, Abhinav Kumar Khorwal and Sujoy Saha are acknowledged for resistivity and magnetic measurements on the PPMS facility at the Central University of Rajasthan. Vikram Singh and Arun Kumar are acknowledged for helping with performing resistivity measurements and synchrotron XRD measurements respectively. Dibyata Rout is acknowledged for helping with heat capacity measurements. SC and SN acknowledge funding from an Air Force Research Laboratory Grant (FA2386-23-1-4046). 

\end{acknowledgments}

\bibliography{Nabib}

\end{document}